\documentstyle[epsf]{mn}
\def\rund#1{\left( #1 \right)}
\def\eck#1{\left\lbrack #1 \right\rbrack}
\def\wave#1{\left\lbrace #1 \right\rbrace}
\def\abs#1{\left\vert #1 \right\vert}
\def\part#1#2{{\partial #1\over\partial #2}}
\def\ave#1{\left\langle #1 \right\rangle}

\begin{document}

\title{The z=2.72 galaxy cB58: a gravitational fold arc lensed by the cluster
MS 1512+36
\thanks{Based on observations made with the NASA/ESA Hubble Space
Telescope, obtained at the Space Telescope Science Institute.}}

\shorttitle{The z=2.72 galaxy cB58: a gravitational fold arc lensed by
the cluster MS 1512+36}

\author[S. Seitz, R.P. Saglia, R. Bender, U. Hopp, P. Belloni \& B. Ziegler]
       {Stella Seitz,$^1$ R.P.  Saglia,$^2$ Ralf Bender,$^2$
	Ulrich Hopp,$^2$ Paola Belloni$^2$ 
        \newauthor 
        and Bodo Ziegler$^{2}$\\
	$^1$Max-Planck-Institut f\"ur Astrophysik, P.O. Box 1523,
	D--85740 Garching, Germany\\
	$^2$Universit\"atssternwarte M\"unchen, Scheinerstr. 1, 
	D--81679 M\"unchen, Germany}

\maketitle

\begin{abstract}
Using HST WFPC2 V- and R-band data of the $z=0.37$ cluster MS1512+36
we show that the $z=2.72$ `protogalaxy' cB58 is not extraordinarily
luminous intrinsically but lensed into a gravitational fold arc by the
cluster.  The arc has a surface brightness weighted axis ratio of
$1:7$, is marginally resolved in width and about $ 3 {\arcsec}$
long. Its counterimage was identified and found to be very compact
($r_{1/2} = 2.4-4.0 h_{50}^{-1}{\rm kpc}$ in a $q_0=0.05$ cosmology).
In addition, we found three further multiple image systems, one with
five and two with three images each. The positions of the multiple
images can be explained by modelling the light deflection caused by
the cluster and the cD-galaxy with elliptical isothermal
potentials. The major axis of the cluster potential approximately
agrees with that of the cD-light and that of the X-ray
isophotes. Since the multiple images are within $\approx 8 \arcsec $
around the cD galaxy, a cluster core radius - cluster velocity
dispersion degeneracy arises. Interpreting the observations
conservatively, the cluster velocity dispersion and the core radius
are limited to $540-670$ km/s and $5\arcsec$ -- $11\arcsec$,
respectively, and the brightness of the unlensed counterimage of cB58
is about $23.9 \pm 0.3$ (R-band), corresponding to a magnification and
extinction corrected restframe-B band absolute magnitude of
$-24.75\pm0.7$ mag.  The redshifts of the sources of the remaining
three multiple image systems are predicted to be similar to that of
cB58 while a strict upper limit of $4$ is set as they are visible in
B-band ground based data.  That part of the source of cB58 which is
mapped into the arc is reconstructed and its magnification is found to
be $\mu_{\rm arc}\ga 50$. This large magnification explains at least
some of the untypical spectroscopic properties of cB58, e.g. that the
star formation rate seems to be high and uniform and to take place in
a large area.
\end{abstract}
\begin{keywords}
Galaxies: fundamental parameters - Galaxies: clusters: individual - 
Cosmology: gravitational lensing,
dark matter
\end{keywords}

\section{Introduction}
\label{sec:1}

The $z=2.72$ galaxy cB58 -- discovered by the (ground-based)
CNOC-survey of cluster redshifts (Carlberg et al. 1996a,b) -- is one
of the brightest ($m_V\approx 20.6$) high-redshift galaxies presently
known. According to Yee et al. (1996, in the following Y96) the galaxy
is a well resolved $3\arcsec \times 2 \arcsec$ disk-like galaxy, and
thus, also the size of the galaxy is surprisingly large for its
redshift. Several strong absorption lines were identified in the
restframe wavelength range of $1000$\AA$\le \lambda \le 2000$\AA,
which are characteristic for a young stellar population.  Using
multi-color photometry Y96 and Ellingson et al. (1996) investigated
its spectral energy distribution in a broader range, i.e. between
$\lambda=5000$\AA $\;$ and $\lambda=21000$ \AA. From SED-models of
Bruzual \& Charlot (1993) they inferred that a substantial fraction of
the stellar component of the galaxy is younger than 10 Myrs, and that
the extinction-corrected and model-dependent star formation rate has
to be of order $4700 M_\odot {\rm yr}^{-1}$. Due to the homogeneous
appearance of the galaxy, the stars have to form uniformly in the
galaxy. Thus, they conclude that cB58 is a galaxy in its initial star
formation stage and call it a `protogalaxy'. A large magnification by
gravitational lensing which would decrease the `true' magnitude, size
and star-formation rate of the galaxy was discarded as unlikely due to
the `resolved, regular and smooth nature of the object'.
 
Since the galaxy cB58 lies only $6 \arcsec$ away from the central
galaxy of the cluster MS 1512+36 at $z=0.373$, Williams \& Lewis
(1997, in the following WL) investigated the possibility that cB58 is
a `normal' $z\approx 3$ star-forming galaxy magnified by the cluster.
According to WL a magnification of $\mu \approx 40$ is sufficient to
decrease the observed non-extinction corrected star-formation rate of
$400 M_\odot {\rm yr}^{-1}$ to that value found by Steidel et
al. (1996a) for $z>3$ galaxies and by Ebbels et al. (1996) for arcs in
galaxy clusters.  The mass distribution of the cluster was modeled as
an isothermal sphere with a core. A velocity dispersion of
$\sigma=1000 { \rm km/s}$ and a core radius of $22\farcs 2$ provides a
(fine-tuned) model for the mass distribution with a large
magnification of $\mu \approx 40$ but a small shape distortion at the
position of cB58, and avoids the prediction of multiple images, which
where not observed from ground.

The velocity dispersion used in WL is in conflict with the measured
value of $(690\pm 100) {\rm km/s}$ by Carlberg et al. (1996a), with a
$90$ percent confidence interval of $602 \le \sigma/{\rm km/s} \le
840$ (Carlberg, private communication). The core radius of $\approx
140 h_{50}^{-1} {\rm kpc} $ exceeds that of the most detailed modeled
and more massive clusters, e.g. A370 (Mellier et al. 1990, Kneib et
al. 1993) or A2218 (Kneib et al. 1994, 1996) by 50 percent and a
factor 3.5, respectively.

In this paper we present two color WFPC2 HST-observations.  In
contrast to the ground based data used by Y96 one can infer from the
high-resolution WFPC2 data that although the cluster is poor in terms
of velocity dispersion and optical richness, it is able to strongly
distort and produce multiple images of high-redshift galaxies, and
that cB58 is a gravitational fold arc.  In section 2, 3 and 4 we
describe the observations, the observed properties of the cluster, of
some of its galaxies, and of all multiple image systems found. The
positions of the multiple images are used for the lens modelling in
section 5, where also limits on the magnification of the counterimage
of cB58 are derived. The magnification of the arc cB58 is estimated in
section 6. Section 7 provides a weak lensing analysis which serves as
a consistency check for the estimated velocity dispersion. The results
are summarized and discussed in section 8.

\section{Observations}
\label{sec:2}

The cluster MS 1512+36 was observed with HST as part of a program to
study the evolution of the Fundamental Plane of elliptical galaxies as
a function of redshift and to constrain the geometry of the Universe
(Bender et al. 1997). A detailed description of the corresponding
results can be found in Saglia et al. (1997).  Using the WFPC2 and the
filters F555 and F675, three orbits were cumulated for each filter for
a total of 6.3 and 5.8 Ksec, respectively.  The exposure time was
split in nine dithered images per filter to increase the resolution of
the final image. Three sets of three images were taken, with
horizontal integer pixel shifts between the three images and vertical
subpixel (1/2 and 1/4 of the WF pixel size) shifts between the sets,
to allow an optimal cosmic ray rejection.  The pipeline reduction was
checked to be adequate. The images with integer pixel shifts were
combined using the cosmic ray rejection IRAF
\footnote{IRAF is distributed by the National Optical Astronomy
Observatories which is operated by the Association of Universities for
Research in Astronomy, Inc. under contract with the National Science
Foundation.} algorithm, rebinned to a pixel size of one quarter of a
WF pixel, aligned, averaged and rebinned to a pixel size of half a WF
pixel, i.e. $0\farcs0498$.  The zero-point photometric calibration was
computed following Holtzman et al. (1995) and found to be in agreement
with ground based photometry of the cluster (Ziegler 1996, 1997).  The
subsequent reduction was performed using MIDAS. The isophote shape
analysis of the central cD galaxy allowed to determine a first
estimate of the position of the major axis of the cluster potential,
using the algorithm of Bender \& M\"ollenhoff (1987), adapted to the
HST resolution. A model for the light distribution of the cD galaxy
was constructed from the isophote shape analysis and subtracted from
the images.

%%%% MAGNITUDES
The magnitudes and colors of the gravitationally lensed galaxies
described below were derived from these frames computing both aperture
photometry with appropriate apertures and annuli for the estimation of
the sky value, and isophote magnitudes. Table 1 gives the magnitudes
and colors inside the 24.78 mag/arcsec$^2$ isophote in the V and 24.71
mag/arcsec$^2$ isophote in the R band, corresponding to the $3-\sigma$
limit above the sky. The errors are computed from the isophote
magnitudes above 2 and $4-\sigma$ above the sky and reproduce the
observed variations in the aperture magnitudes due to photon
statistics of when different galaxy or sky apertures are used.

%%%% DECONVOLUTION
The Lucy-Richardson deconvolution algorithm (Lucy 1974) as implemented
in MIDAS was applied to enhance the resolution across the
gravitational arc cB58. Twenty iterations were performed using the PSF
generated by TINYTIM.

With the exception of the cD galaxy, all galaxy shapes were estimated 
using SExtractor (Bertin \& Arnouts 1996).
\section{The cluster and the cD-galaxy}
\label{sec:3}
Fig. 1 shows the core of the cluster MS1512+36 using coadded V- and
R-data. The cluster is dominated optically by its cD galaxy (center of
Fig. 1). The measured velocity dispersion of the cD galaxy equals
$260\pm 20 {\rm km/s}$, and increases to $\sigma_{\rm cD}=286 \pm
20 {\rm km/s}$ after aperture correction (Ziegler \& Bender 1997).
The results of the isophote shape analysis of the cD galaxy are shown
in Fig. 2.

Note that the position angle and eccentricity of the light from the
halo of the cD are not constant but both increase in the outer parts
of the cD. From Fig. 2 we derive an axis ratio and major axis of the
cD of $r=b/a=0.7$ ($r=0.6$) and $\phi=10 \degr$ ($\phi=6\degr$) at a
distance of $4\arcsec$ ($7.5\arcsec$) from its center (angles are
counted with respect to the x-axis of the WFPC 3-chip).  The major
axis of the cD galaxy and that of the cluster potential -- as inferred
from X-ray photons -- are roughly in agreement (compare with the X-ray
map in Hamana et al., 1997).  Similar to the optical data the X-ray
map shows also a twist of the isophotes and an ellipticity change of
the X-ray contours.

There is a face-on blue spiral galaxy with $m_V=21.08$ and
$m_{V-R}=0.48 $ at a distance of 4 arcseconds from the cD galaxy.
Absorption of the light from the halo of the cD near the spiral arms
suggests that the spiral is in front of the cD.  With only one color
it is quite difficult to estimate its photometric redshift. However,
the irregular morphology of the galaxy, characterized by very bright
and numerous H II star forming regions allows us to put some
constraints on its spectral type and star formation history.  We
modeled the galaxy's stellar population using the GISSEL library for
solar metallicity and Salpeter IMF (Bruzual \& Charlot 1997),
assuming an exponentially decreasing star formation rate
$\psi(t)=\tau^{-1}\exp(-t/\tau)$ with different time scales
($7<\tau<10$ Gyr), and allowing for different return fractions for
gas recycling (up to 40\%).  We have assigned an age of 10 Gyr to the
galaxy and considered the evolution of its spectral energy
distribution for a set of cosmological parameters ($H_o= 50,75 {\rm
km /s/Mpc}^{-1}$, $q_0=0.01,0.1$).  There are many uncertain
parameters in these models, like the IMF, metallicity, dust etc.
However, all galaxy models we considered indicate that the observed
$V-R=0.48 $ can be matched either at z $\leq 0.35$ or at z $\approx$
0.9.  The exponential scale length of $1\farcs 1 - 1\farcs 3$ clearly
excludes the second hypothesis and thus the most probable redshift is
$0.25 < z < 0.37\equiv z_{\rm cl}$.  For this redshift range the R
magnitude corresponds approximately to the rest frame V magnitude and
the luminosity of the galaxy is 0.8$L_* \leq L \leq 1.8 L_* $.

\section{Multiple Images}
\label{sec:4}
Although the cluster is optically poor and dynamically weak,
there are several strong and weak lensing signatures visible:
\label{fig:3.1}
\begin{figure*}
\begin{center}
  \epsfxsize=\hsize\epsffile{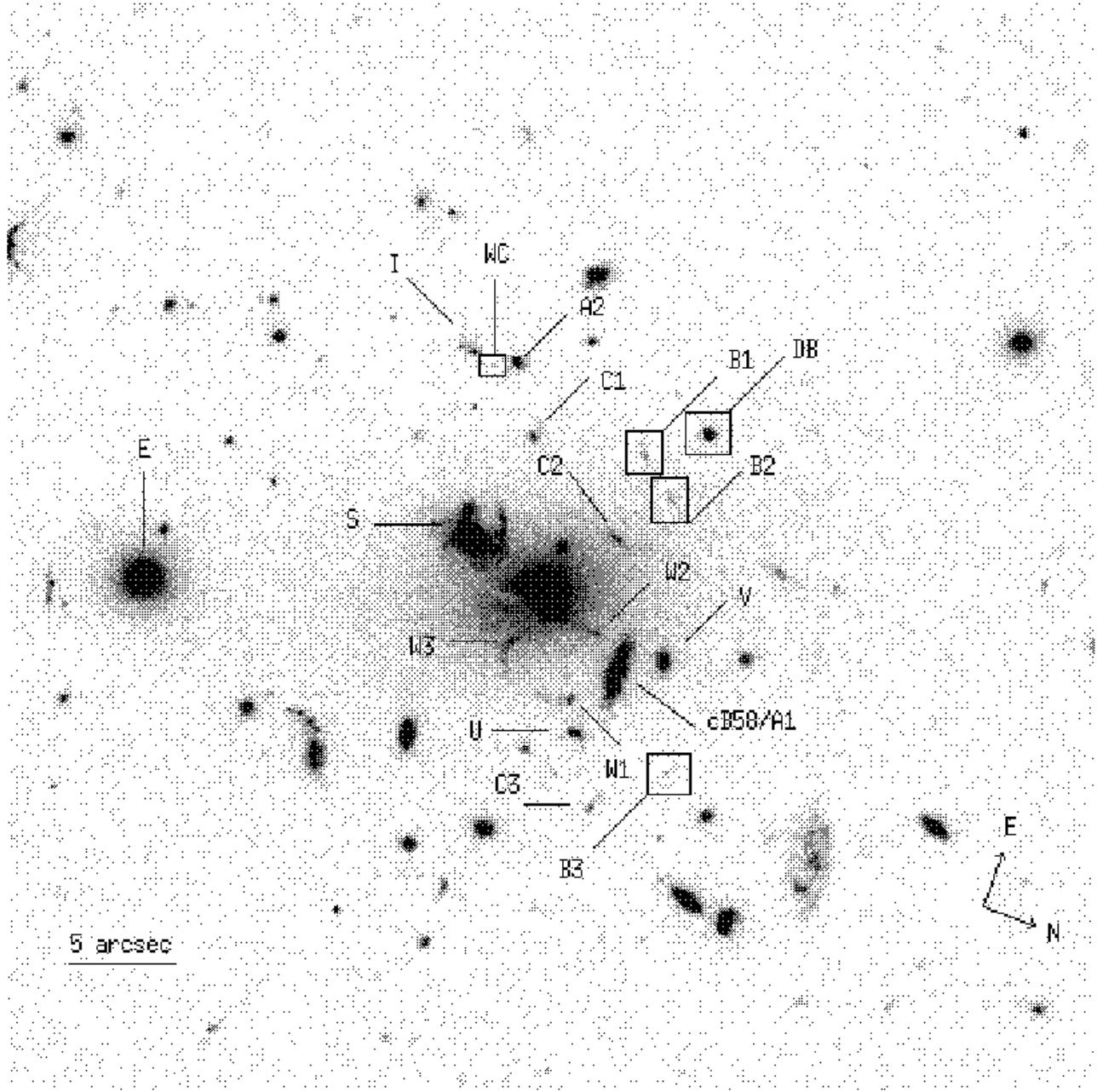}
\end{center}
\caption{Image of the core of the cluster MS1512+36 using coadded V-
and R-data. Close to the cD galaxy in the center is a face on
spiral S to its left. The galaxy cB58 is on the opposite side at
$5\arcsec$ distance with an inclination angle of $71.6 \degr$ relative
to the x-axis. The compact bright object A2 at $(r,\phi) \approx (11\arcsec,
97 \degr)$ is the counterimage of the gravitational arc cB58. Left to
the upper and lower end of cB58 are two shrimp-like objects (W1 \& W2)
with their heads pointing towards cB58. As argued in the text, they
are also lensed, with a counterimage WC left of A2, and a fourth image
at W3. Near the upper right diagonal of the field there is the
galaxy pair B1 \& B2 (see also Fig. 4). The counterimage candidate B3
with polar coordinates $(r,\phi)\approx (10 \arcsec, -56 \degr)$
relative to the cD is marginally visible in this Figure (see also
Fig. 4). The three galaxies C1, C2 and C3 are most likely also multiply
imaged galaxies.
}
\end{figure*}

\label{fig:2.1}
\begin{figure}
\begin{center}
  \epsfxsize=\hsize\epsffile{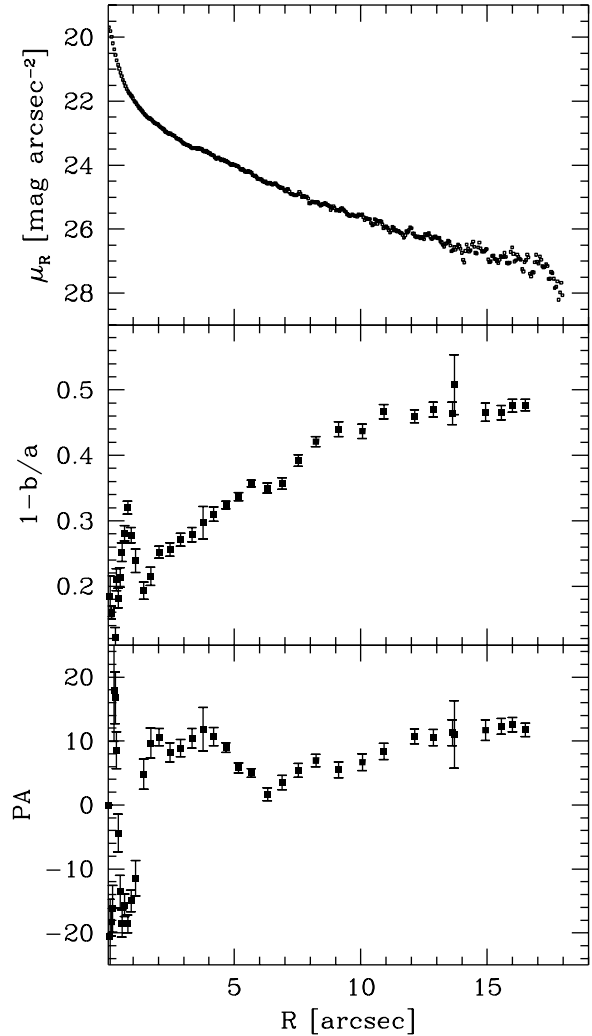}
\end{center}
\caption{The isophote shape analysis of the cD galaxy. The top panel
gives the circularly averaged surface brightness profile in the R
band. The panel in the middle shows the ellipticity profile as a
function of the circularized distance from the centre $R=\sqrt{ab}$. 
The panel at the bottom shows the position angle profile in degrees as
a function of $R$.}
\end{figure}

\begin{figure}
\begin{center}
  \epsfxsize=\hsize\epsffile{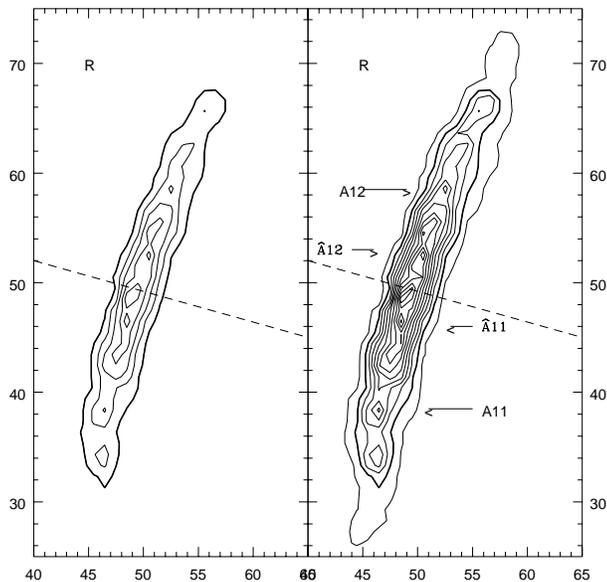}
\end{center}
\caption{R-band counts for the galaxy cB58 obtained from the
 Lucy-Richardson deconvolved data. One unit in the horizontal and vertical
 direction equals one pixel of size $\approx 0\farcs 05$.  Contours
 are spaced by 10 (left) and 5 (right) counts per pixel; the thick
 line corresponds to 10 counts per pixel.  The knots marked by A11
 \& A12, \^A11 \& \^A12 are used for the lens modelling later on.
}
\end{figure}

i)
As described by Y96, the galaxy cB58 (also denoted by A1 in the
following) is at $5\arcsec$ distance from the center of the cD; its V-
and R-band magnitudes are $m_R=20.29$ and $m_V=20.64$, and its
major-axis position angle is $\phi_{A1}=71.6 \degr $. The galaxy is
more elongated than visible from the ground (compare with Fig. 2 in
Y96).  Its light distribution is slightly curved, with the center of
curvature not coinciding with the position of the cD, but pointing
towards the outskirts of the cluster.  Fig. 3 shows the
Lucy-Richardson deconvolved light distribution of cB58 in the
R-band. The local background and rms-noise are 1.2 and 0.4 counts per
pixel, respectively; hence the limiting contour of 5 counts per pixels
is 9.5 sigma above the background and the light distribution inside
this contour is hardly affected by background noise.  The mirror
symmetry of the light distribution (the symmetry axis is sketched by
the dashed line) indicates that cB58 is a merged image pair of a
gravitationally lensed source. The gravitational arc is very elongated
and only marginally resolved in width: perpendicular to the major axis
the flux distribution of the galaxy is confined to 4-5 (dithered)
pixels, and the flux increases steeply from the boundary towards the
major axis. At the contour level of 5 counts per pixel, the extent of
the galaxy parallel to the major axis is about 50 pixels. The
SExtractor axis ratio obtained from the (8) isophote-weighted second
moments of the light distribution (within the same contour as limiting
isophote) is about 7. The unweigthed ratio of the outermost isophote
is about $1:10$.  This axis ratio also places a lower limit on the
magnification of the arc, together with its unresolved width and
assuming that the source of A1 is spherical.  The area enclosed by
pixels with a surface brightness larger than 4 (5.5,10) counts per
pixel is equal to 258 (210,138) pixels.
\label{fig:3.3}
\begin{figure}
\begin{center}
  \epsfxsize=\hsize\epsffile{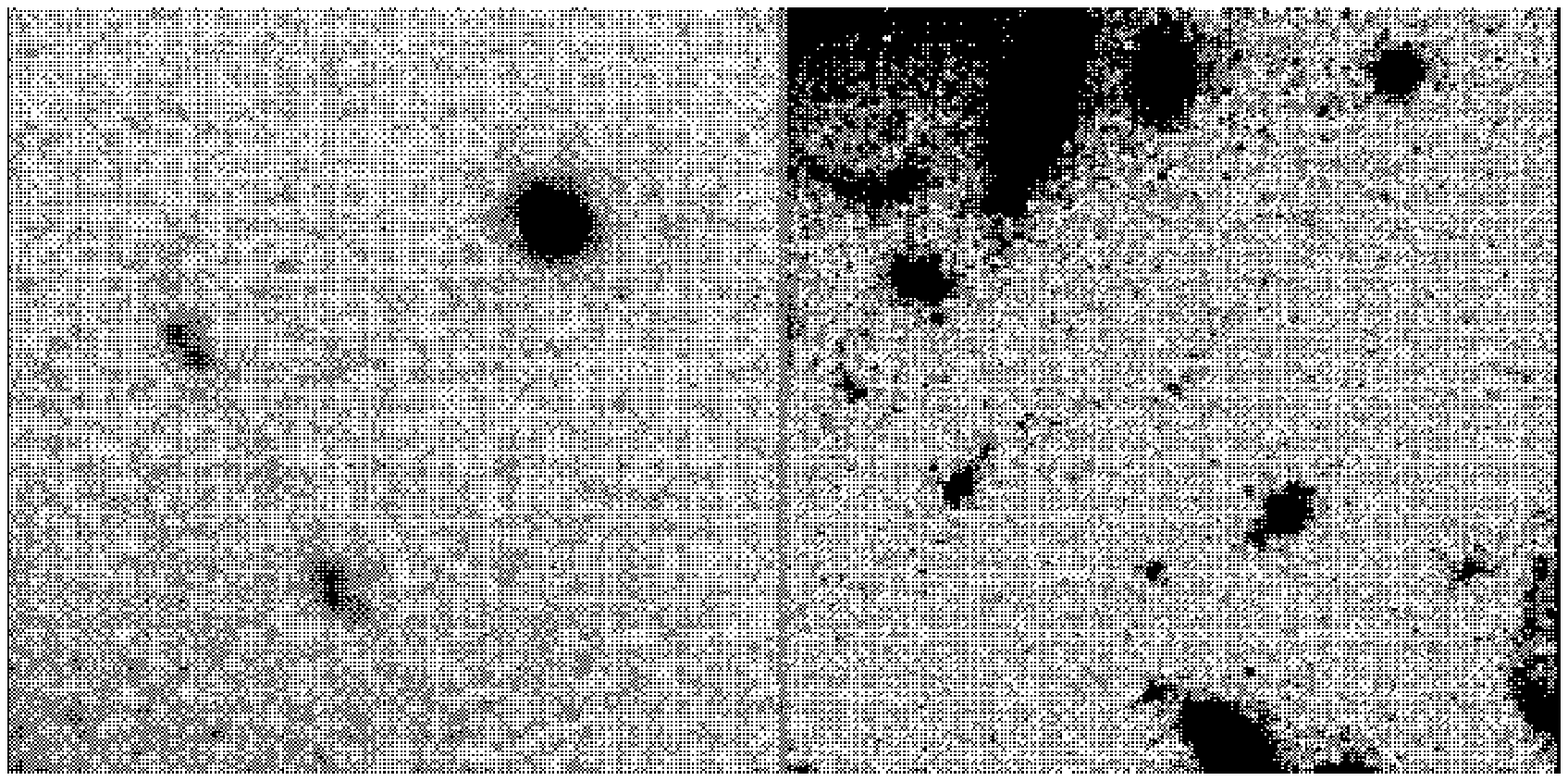}
\end{center}
\caption{The left panel shows the pair of galaxies B1 (upper one) and
B2 (lower one) which is at $8\arcsec$ distance from the cD
galaxy. Their morphology and colors are identical and a diffuse
emission is connecting them. The elliptical galaxy DB next to B1 and
B2 will slightly perturb the tangential critical line caused by the
smooth cluster potential.  In the center of the right panel the
counterimage candidate B3 is visible. It is barely resolved, but shows
an additional diffuse emission which could correspond to the `arc'
connecting B1 and B2.}
\end{figure}

ii) There are several indications that the faint galaxies (B1, B2 in
the following) at $8\arcsec$ distance from the cD-galaxy belong to a
multiple image system: The galaxies have similar morphologies (see
Fig. 4) and the same colors within the quoted errors (see Table 1).
Third, there is a diffuse emission connecting the two galaxies, as
expected when a faint part of the source lies on a caustic and thus is
mapped into a gravitational arc, whereas the major part of the source
is inside that caustic and is mapped into two images separated by the
corresponding critical line. From the $V-R\approx 0.2$ color of B1 and
B2 and from the reasonable assumption of their morphological (spiral)
type we estimate $z_B>1.5$ as a lower limit for their redshifts with
the GISSEL models as above. The upper limit is $z_B<5$ since otherwise
the galaxies restframe Lyman-limit would drop out of the observers
V-band. The elliptical galaxy (DB) $3\farcs 2$ apart from $B1$
coincides in color (see Table 1) with that of spectroscopically
confirmed cluster members (e.g. the elliptical E and the cD galaxy;
for details about spectroscopy of cluster members see Carlberg et al.
1996a\&b, Ziegler \& Bender 1997).  The $D_n$-$\sigma$-relation
(Dressler et al. 1987) yields an estimate of $84\pm15 {\rm km/s}$ for
its velocity dispersion. Hence, the galaxy DB is separated by about 22
Einstein radii from B1 and its light deflection on B1 and B2 could be
neglected in the field or in regions where the cluster is weak. Nearby
a critical line, however, the small shear induced by DB is sufficient
to perturb it locally and thus to change the magnification and
(slightly) the positions of B1 and B2. Therefore, the galaxy DB will
be taken into account in the quantitative lens modelling below.

The cluster potential determines the global properties of the
tangential critical line of the lens systems (like the enclosed area
and its approximate shape), whereas the massive cD galaxy and small
galaxies near the critical line can perturb it only locally.  Although
the redshift of the galaxies B1 and B2 is not known, one can conclude
that the cluster potential must be quite asymmetric already from the
positions of the arc A1 and the galaxies B1 and B2: On the one hand,
one could account for the fact that the direction and curvature of the
arc is not in agreement with a spherically symmetric model by
assigning sufficient lensing strength to the small galaxy V to the
right of A1, perturbing the critical line and adding additional shear.
On the other hand, if the cluster is spherically symmetric and has a
small core radius as usually found for non-massive clusters (Mellier
et al. 1993), the position of A1 indicates the Einstein radius of a
$z=2.7$ source, and the separation of the B1 and B2 system from the cD
center (which is assumed to agree with the cluster center) measures
the Einstein radius for the B1 and B2 source. Under the assumption of
spherical symmetry the ratio of these two Einstein radii is $1.4 -1.6$
and it then describes the angular diameter distance from the cluster
to the source of B1 and B2 in units of the angular diameter distance
from the cluster to cB58 at $z=2.7$. Since cB58 is at a high redshift
already, such a high ratio can not be achieved for reasonable
cosmologies, eg. $\Omega+\lambda =1$ with $\lambda \le 0.7$. If the
core radius is non-negligible relative to the Einstein radius and is
of the order of 6 to 7 arcsec as found by Hamana et al. (1997), the
ratio of the Einstein radii drops to about $1.2$ which is still too
large to be accounted for by spherical symmetry.

The parameters of a lens system can be determined most accurately if
the corresponding counterimages of arcs or double images on the
opposite side of the cluster are found, because this determines the
enclosed mass most strongly (see eg. Mellier, Fort \& Kneib 1993 for
MS 2137-23, Kneib et al. 1994, 1996 for Abell 2218).  Modelling the
cluster as an elliptical non-singular isothermal potential, and the cD
as an elliptical singular isothermal potential and fitting the
positions of the arc A1 and the pair B1 and B2 (with unknown redshift)
we can predict the positions of the counterimage of A1 (denoted by A2)
and that of B1 and B2, denoted by B3. Comparing to the observations we
identify a galaxy as the counterimage A2 and a candidate for the
counterimage B3.
\label{fig:3.4}
\begin{figure}
\begin{center}
\epsfxsize=\hsize\epsffile{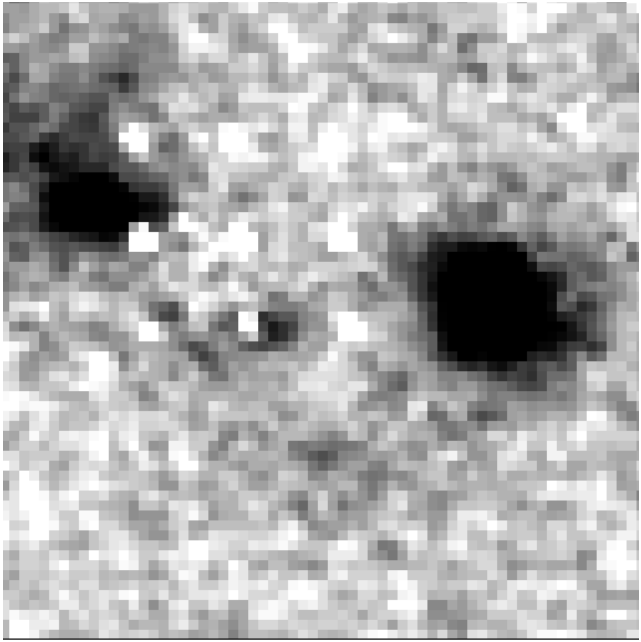}
\end{center}
\caption{The right object in this part of the $V+R$-coadded image is
the counterimage A2 of the gravitational arc cB58; the left object (I)
is considerably bluer than A2. The grid of nine white pixels is caused
by one bad pixel and reflects the nine different dither positions in
the coadded $V+R$-exposures. The small object (WC) between A2 and I
has a morphology similar to the `shrimps' W1 and W2 visible in Fig. 1
and Fig. 7. We consider this as the counterimage candidate of W1 and
W2.}
\end{figure}
\label{fig:3.5}
\begin{figure}
\begin{center}
\epsfxsize=\hsize\epsffile{counts_R_A2.teps}
\end{center}
\caption{The upper part of this figure shows R-band contours of counts
per pixel within the approximately same region as seen in Fig. 5. The
thin curves represent low signal-to-noise contours with 1.35, 1.45,
1.55, 1.65 and 1.75 counts per pixel; the object WC seen in Fig.5 can
barely be recognized. The thick contours start at 2 counts per pixel
and increase in steps of two, thus flagging the high signal-to-noise
objects A2 and I. The surface brightness of A2 increases steeply
towards its center in the R-band (lower left) and the V-band (lower
right). Here, the contours also start at 2 counts per pixel and
increase in steps of two; the thick contour is that of 10 counts per
pixel.
}
\end{figure}

iii) 
The galaxy A2 is the compact object at the upper boundary of Fig. 1
and, more expanded, the bright object at the right of Fig. 5. Its magnitudes
are $m_{R}=22.83$ and $m_V=23.23$; it is the only object in that
region where the counterimage is expected and whose color agrees with that
of A1 (see Table 1). The galaxy to the left of A2 (denoted by I) is too blue
to be the counterimage. The contours for the counts per pixel in the
R- and V-band can be seen in the lower left and lower right part of
Fig. 6. The high surface brightness core of A2 is unresolved. In the
R-band, the surface brightness of 4 (5.5, 10) counts per pixel is only
exceeded within 27 (16, 3) pixels. A comparison of the area enclosed by
the same contours in A1 shows that the arc-area is $\ge 46$ 
(approx. 13, approx. 10) times that of the counterimage for contours of 10
(5.5,4) counts per pixel.
From the magnitudes of A1 and A2 we obtain a
magnification ratio of A1 to A2 of $\mu_{A1/A2} \approx 11$.  Hence we
conclude that the high surface brightness core of the source of A1 and
A2 lies on the caustic and is magnified most strongly, whereas the
other remaining regions of the source are magnified only moderately, and
parts of it are only singly imaged (into A2). 

iv) 
The candidate B3 for the counterimage of B1 and B2 is the central
object in the right panel of Fig. 4 which was obtained using coadded
$V+R$ data. Since its flux is not much above the noise level in R, the
magnitude there can be measured only with a large error; this is less
severe in the V-band; the color of B3 agrees within the large errors
with that of B1 and B2. The galaxy in the center of Fig. 4 (right
panel) is the only
object above the $5\sigma$ detection limit of $V= 28.2$ whose
position and color is compatible with being the counterimage
B3. Although there is a galaxy nearby B3 as bright as DB (but bluer)
for the case of B1 and B2, the additional light deflection caused by
this galaxy will not be considered, since the cluster is non-critical
at B3.

\label{fig:3.6}
\begin{figure}
\begin{center}
  \epsfxsize=\hsize\epsffile{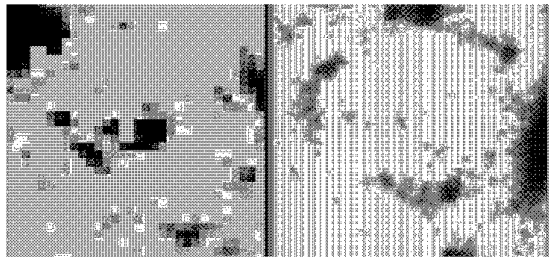}
\end{center}
\caption{In the left panel one can see  the counterimage candidate
WC; although the signal-to-noise for this object is low (even in the
coadded $V+R$ data), one can identify the same shrimp morphology (`head' to
the right and curved `tail' to left) as it is seen in the objects W1
(lowest object), W2 (right object) and $W3$ (left object) in the right
panel, where a model for the emission of the cD galaxy was subtracted.}
\end{figure}

v) To the left of the upper and lower end of cB58 (see Fig. 1) are two
shrimp-like objects (the lower (upper) one is denoted by W1 (W2))
pointing with their `head' towards cB58. These two images have
different parity and they are on opposite sides of the critical line
defined by the flux distribution of the arc (see Fig. 3). Hence, a
lensing interpretation is most natural, and it suggests that the
source redshift is similar to that of cB58. Simple lens models
(using the positions of A1 and A2, B1 and B2 to determine the parameters
of the cD and the cluster) indeed {\it predict} that W1 and W2 have a
common source and a faint counterimage left to A2 if they are at the
same redshift as A1. They also predict a fourth image (W3 in the
following) to the left of
W1 and W2 right at that point of blue emission which makes the W-system
to appear as a circular structure. The position of the fifth image
depends more on the details of the model: eg., it can merge with the
fourth image causing the more extended `head' of W3, or it could be a
fainter image near the center of the cluster.  We can also identify
the predicted counterimage of the W-system with a noisy emission
between the I and A2-galaxies in Fig. 5 (WC in the following). 
More details of the W-system
are given in Fig. 7 (using $V+R$ data). The left panel of this figure
shows that the object WC is also `shrimp-like' and has the same parity
as W1. Since the fluxes of W1, 2 and 3 in the R-band are considerably
affected by the red light of the cD galaxy, color comparisons are
difficult. As described before, a model for the light distribution of
the cD galaxy was constructed from the isophote shape analysis and
subtracted from the images. But especially near W3 the color estimates
may be affected by dust absorption due to the nearby spiral. For
W1, 2 and WC the colors agree within the error bars, which do not
include systematic errors
due to insufficient subtracted cD light. We conclude that the `true'
color of the W-system is likely to be that of W1 or WC, whereas W2 and
W3 may be affected by inaccurate subtraction of the cD light.

\hfill\break
\label{fig:8}
\begin{figure}
\begin{center}
  \epsfxsize=\hsize\epsffile{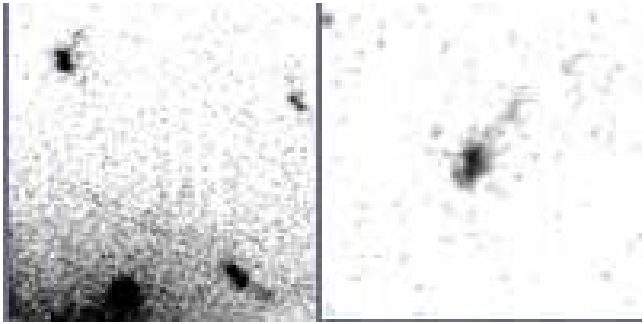}
\end{center}
\caption{In the upper left and lower right corner of the left part of
this figure C1 and C2 are visible. The suggested counterimage C3
(right part of the figure) coincides in morphology with C1 and C2: all
objects consist of a `head' and a `tail' inclined to the major axis of
the head. The parity of C1 and C2 is reversed. The relative
orientation of the tail is predicted by lens models.
}
\end{figure}
vi) The three galaxies C1 to C3 in Fig. 1 are most likely multiply
lensed as well. C1 and C2 are on the opposite sides of the critical
line for a $z\approx 2-3$ galaxy as it can be seen from an
extrapolation of the
critical line passing through B1 and B2 to the left.  All of the three
galaxies have a `head' and a `tail' inclined to the major axis of the
head (see Fig. 8). C2 has opposite parity as C1 and C3.  The colors of
C1 and C3 do not support the lensing hypothesis (Table 1) because they
only marginally agree within their 1-$\sigma$ errors. The ground-based
data in the B-band (Gioia \& Luppino, 1994) provided by G. Luppino
neither contradict nor confirm the lensing hypothesis further;
although C1, C2 and C3 are all visible in the data, the exposure is not
deep enough to improve the accuracy on the color determination. As
the C-galaxies as well as B1 and B2 are visible in the
B-band we can estimate the  upper limit of their redshifts 
to about $z_{max}=4$.

Despite the slight inconsistency of the colors of  C1 and  C3 we
consider C1, C2 and C3 as 
a multiple image  system:
for every lens model which reasonably fits the A, B and W
system we {\it robustly predict} that C1 to C3 have a common source if their
redshift is marginally larger than that of cB58. Additionally, the
directions of the tails relative to the major axis of the heads are
also predicted as they are observed. Accidentally it may happen that three
sources are along the line of sight to three possible multiple image
positions, but we consider it unlikely, since not only their
relative morphology but also their relative orientation agrees with
that predicted by gravitational lensing.

% table 1
%
\begin{table}
\caption{R- and V-magnitudes of objects as defined in the text (see
Fig. 1).}
\label{tab:2.1}
\begin{center}
\begin{tabular}{|l|r|l|l|l|r|}
\hline
Object & $m_R$ & $\Delta m_R$ & $m_V$ & $\Delta m_V$ & $V-R$\\
\hline
A1    & $20.29$  & $+0.01,-0.03$  & $20.64$  & $+0.02,-0.02$  & $0.41$  \\
A2    & $22.83$  & $+0.04,-0.07$  & $23.23$  & $+0.05,-0.06$  & $0.40$  \\
\hline
B1    & $25.77$  & $+0.27,-0.31$  & $26.08$  & $+0.16,-0.37$  & $0.31$  \\
B2    & $25.69$  & $+0.29,-0.25$  & $25.90$  & $+0.24,-0.26$  & $0.21$  \\
B3    & $28.42$  & $+1.64,-1.30$  & $27.58$  & $+0.45,-0.71$  & $-0.84$ \\
\hline
I     & $24.08$  & $+0.19,-0.18$  & $24.22$  & $+0.18,-0.22$  & $0.27$  \\
%\hline
%C1$^0$    & $24.60$  & $+0.20,-0.23$  & $24.66$  & $+0.20,-0.22$  & $0.06$  \\
%C2$^0$    & $24.98$  & $+0.21,-0.21$  & $24.98$  & $+0.19,-0.19$  & $0.00$  \\
%C3$^0$    & $24.95$  & $+0.28,-0.25$  & $25.58$  & $+0.17,-0.33$  & $0.63$  \\
\hline
C1    & $24.60$  & $+0.20,-0.23$  & $24.66$  & $+0.20,-0.22$  & $0.06$  \\
C2    & $24.98$  & $+0.21,-0.21$  & $24.98$  & $+0.19,-0.19$  & $0.00$  \\
C3    & $24.95$  & $+0.28,-0.25$  & $25.58$  & $+0.17,-0.33$  & $0.63$ \\
%\hline
%W1$^0$    & $24.42$  & $+0.28,-0.27$  & $24.78$  & $+0.34,-0.38$  & $0.36$  \\
%W2$^0$    & $24.82$  & $+0.27,-0.26$  & $25.61$  & $+0.46,-0.49$  & $0.79$ \\
%W3$^0$    & $24.73$  & $+0.30,-0.32$  & $25.02$  & $+0.44,-0.39$  & $0.29$  \\
\hline
W1    & $24.42$  & $+0.28,-0.27$  & $24.78$  & $+0.34,-0.38$  & $0.36$  \\
W2    & $25.82$  & $+0.27,-0.26$  & $25.61$  & $+0.46,-0.49$  & $0.79$ \\
W3    & $24.73$  & $+0.30,-0.32$  & $24.02$  & $+0.44,-0.39$  & $0.29$  \\
WC    & $26.89$  & $+1.12,-0.55$  & $27.04$  & $+0.68,-0.60$  & $0.15$  \\
\hline
U     & $23.66$  & $+0.09,-0.13$  & $24.16$  & $+0.10,-0.12$  & $0.50$  \\
V     & $22.96$  & $+0.06,-0.09$  & $23.30$  & $+0.14,-0.10$  & $0.34$  \\
\hline
DB    & $22.64$  & $+0.03,-0.06$  & $23.80$  & $+0.04,-0.08$  & $1.16$  \\
\hline
S     & $20.60$  & $+0.03,-0.04$  & $21.08$ & $+0.04,-0.06$    & $0.48$ \\
\hline
\end{tabular}
\end{center}
\end{table}

\section{Modelling the lens system }
\label{sec:5}
\subsection{The deflection potential of the 
cluster and the galaxies}
\label{sec:5.1}
We model the deflection potential of the cluster, the cD-galaxy and of
other individual galaxies by an elliptical non-singular isothermal
deflection potential with a velocity dispersion $\sigma$, a core
radius $\zeta$, a major axis orientation angle $\phi$ and an
ellipticity parameter $\epsilon$ [in units of ${\rm km/s}$, (dithered)
pixels, and degrees with respect to the x-axis],
\begin{equation}
\psi(\tilde x,\tilde y)=\psi_0\sqrt{1 + q_\psi  \; \rund{\tilde x\over \zeta}^2
+ {1\over q_\psi} \;  \rund{\tilde y\over \zeta}^2}  \; .
\label{eq:2.2}
\end{equation}
Here, $\tilde x$ and $\tilde y$ denote pixel positions with respect to
the center of mass, the $\tilde x$-axis is parallel to the major axis
of the potential and $q_\psi=:
(1-\epsilon)/(1+\epsilon)$ is the axis 
 ratio of  equipotential contours.  The normalization
$\psi_0=\theta_{\rm E} \; \zeta $ depends on the Einstein radius
\begin{equation}
\theta_{\rm E}= {4\pi} \rund{\sigma  \over c}^2 \; {D_{\rm ds}\over
D_{\rm s}} \; ,
\label{eq:2.3}
\end{equation}
where $c$ is the speed of light while the source redshift and the
cosmological 
parameters enter in the
angular diameter distances from the cluster to the source ($D_{\rm ds}$)
and from the observer to the source ($D_{\rm s}$). At redshift $z$ the
Einstein angle becomes 
\begin{equation}
\theta_{\rm E} \approx \rund{\sigma\over 49 {\rm km/s}}^2 
\; {D_{\rm ds}(z)\over D_{\rm s}(z)} 
\eck{D_{\rm ds}(cB58)\over D_{\rm s} (cB58)}^{-1} \; {\rm pix}
\end{equation}
for an  Einstein-de
Sitter universe and increases by a factor of $\approx 1.4$ for a flat
universe with $\Omega=0.3$. Any value of  velocity dispersions quoted
below assumes an Einstein-de Sitter cosmology.

The deflection angle
$\alpha$, surface mass density $\kappa$, shear $\gamma=\gamma_1+i
\gamma_2$ and magnification $\mu$ of a point source can be obtained
as first and second order derivatives of the
deflection potential with respect to the angular coordinates $\tilde
x$ and $\tilde y$ (Schneider, Ehlers \& Falco 1992),
\begin{equation}
\alpha_i =  \psi_{,i} \; , \;
\kappa = {1\over 2} \rund{\psi_{,22} + \psi_{,11}}\; , \;
\end{equation}
\begin{equation}
\gamma_1 = {1\over 2} \rund{\psi_{,22} - \psi_{,11}}\;  , \;
\gamma_2 = -\psi_{,12}\;  , \;
\end{equation}
\begin{equation}
\mu^{-1} =(1-\kappa)^2 -\abs{\gamma}^2   \; .
\label{eq:2.4}
\end{equation}
With $C:= \psi/\psi_0$, $Q:=(q_\psi +1/q_\psi)$ and
$\bar Q:=(1/q_\psi-q_\psi)$ these functions become
\begin{equation}
\kappa= {1\over 2 C^3 }\; 
\wave{ Q + {\tilde x^2 + \tilde y^2 \over \zeta^2 }} 
  {\theta_{\rm E} \over \zeta}
\; ,
\end{equation}
\begin{equation}
\gamma_1={1\over 2C^3} \;
\wave{ \bar Q + {\tilde x^2 - \tilde y^2 \over \zeta^2 }} 
  {\theta_{\rm E} \over \zeta}\;  , \;
\end{equation}
\begin{equation}
\gamma_2={1\over 2C^3} \;
\wave{ {2 \tilde x \tilde y \over \zeta^2}}
  {\theta_{\rm E} \over \zeta}
\; .
\end{equation}
The mass distribution has 
 the following properties:
\begin{itemize}
\item[1.] 
It simplifies to the mass profile of a singular isothermal sphere,
 $\kappa \approx {\theta_{\rm E} \over 2 \sqrt{\tilde x^2 +\tilde
 y^2}}$, if the mass distribution is spherically symmetric ($q_\psi=1$)
 and if positions with $\tilde x^2 + \tilde y^2\ \gg \zeta^2 $ are
 considered.
\item[2.]
For an isopotential axis ratio $q_\psi$ the mass within an {\it
ellipse}  of
the same axis ratio $q_\psi$ and area $\pi r^2$ equals
\begin{equation}
M_r(q_\psi)= {\pi \theta_{\rm E} \zeta } \wave{
{Q-1+ {Q\over 2} \rund{r\over \zeta}^2 
\over \sqrt{1+\rund{r\over \zeta}^2 } }  +1-Q }
\; .
\end{equation}
Hence, for an elliptical
potential the ratio of the mass within an ellipse of axis ratio
$q_\psi$ and area $\pi r^2$ to the mass $M_r^0:=M_r(q_\psi=1)$
within a circle of radius $r$ and a circular potential is given by
\begin{equation}
{M_r(q_\psi)\over M_r^0}=
{Q-1 +{Q\over 2} \rund{r\over \zeta}^2 +
\sqrt{1+\rund{r\over \zeta}^2} \eck{1-Q} 
\over
1 +  \rund{r\over \zeta}^2 
- \sqrt{1+\rund{r \over \zeta}^2} }
\; ;
\end{equation}
this ratio equals one at $r=0$, increases monotonically with
increasing $r$ and becomes
\begin {equation}
\lim_{\rund{r\over \zeta} \to \infty} 
{M_r(q_\psi)\over M_r^0}= {1\over 2} Q = \eck{q_\psi +{1\over
q_\psi }}{1\over 2} \ga 1
\; 
\end{equation}
for $r^2 \gg \zeta^2$.  Thus models with the same velocity dispersion
$\sigma$ but different $\epsilon$ have  different isodensity {\it
shapes} but the same `mean' mass density (measured within ellipses
with an axis ratio given by the potential). We want to point out that the
`velocity dispersion' $\sigma$ derived from the amplitude
$\psi_0=\theta_{\rm E} \zeta$  has a straightforward relation to the
observed  velocity dispersion of cluster galaxies only in the
spherical symmetric case. Nevertheless we will express the estimated
amplitude in terms of velocity dispersion of the cluster later on.
\item[3.]
For small eccentricities of the cluster potential $\epsilon _{\rm cl}
< 0.2 $ the isodensity contours are roughly elliptical and the axis
ratio
\begin{equation}
q_\kappa=:{1-\hat \epsilon_{\rm
cl}\over 1+\hat \epsilon_{\rm cl}}
\end{equation}
of the isodensity contours is related  to the isopotential
contours according to $\hat \epsilon_{\rm cl} \approx 3 \;
\epsilon_{\rm cl}$.
\end{itemize}

The lens models investigated below are described by the following
parameters:
\begin{itemize}
\item[1.] 
The orientation $\phi_{\rm cl}$, ellipticity $\epsilon _{\rm cl}$,
velocity dispersion $\sigma_{\rm cl}$ and the core radius $\zeta_{\rm
cl}$ of the {\it cluster} will be treated as  free parameters. The
cluster center is assumed to coincide with that of the cD-galaxy.
\item[2.]
The ellipticity and orientation of the {\it cD galaxy} are either
treated as free parameters, are assumed to be equal to that of the
cluster or inferred from the light. In the second case the assumption
is motivated by earlier investigations (eg., Mellier et al. 1993)
finding that the major axis of clusters dominated by a single mass
concentration is aligned with that of the central galaxy.  In the
third case we assume that the surface brightness distribution of the
extended diffuse emission of the cD traces its surface mass
density. Using equation (13) we obtain $\epsilon_{\rm cD}=0.06-0.08$
and $\phi_{\rm cD}=5\degr - 10\degr$. The core radius is assumed to be
zero, whereas the velocity dispersion $\sigma_{\rm cD}$ of the
cD-galaxy is a free parameter.
\item[3.]
Additional {\it galaxies} are treated as spherical singular systems
($\epsilon=0$, $\zeta=0$) with their velocity dispersion as free
parameters.
\end{itemize}
\subsection{The lens models}
\label{sec:5.2}
Let $\vec p = (p_1,..,p_f) $ denote the free parameters of the lens
model, $K$ the number of multiply lensed sources, $I(k), k =1,..,K$
the number of images for each of the $k$ sources; further, let $\vec
\theta_{i_k}$ denote the position of the $i^{th}$ image ($1 \le i_k
\le I(k)$ ) of the $k^{th}$ source.  For each image position the
deflection angle $\vec \alpha(\vec \theta_{i_k};\vec p)$ is calculated
and a source position 
$\vec \beta_{i_k}(\vec p):=\vec \beta (\theta_{i_k};\vec
p)= \vec \theta_{i_k} -\vec \alpha(\vec \theta_{i_k};\vec p)$ 
is predicted. The best fitting
model is obtained by minimizing
\begin {equation}
E:= \sum_{k=1}^{K} W(k) \sum_{i_k,i^\prime_k=1}^{I(k)}
w(i_k)w(i^\prime_k) 
\abs{\vec \beta_{i_k}-\vec \beta_{i^{\prime}_k}}^2
+ f( P)
\end {equation}
with respect to the free parameters of the model using the {\tt powell}
routine described by Press et al. (1992).  $W(k)$ and
$w(i_k)$ are weights equal to one or zero, and thus they determine
which of the multiply imaged sources are considered ($W(k)$), and
which of their images ($w(i_k)$) are taken into account for the model
fitting.  The additive function $f(P)$ is used to take into account
parities $P$ of images. (The photometry is too inaccurate to include
flux ratios.)

 The fold arc contains much more information than the position of its
 center of light: the critical line passes through it and each pixel
 must have a corresponding one with the same surface brightness on the
 opposite side of the critical line, and these pixel pairs will have a
 common source in the ideal case. The information contained in the
 light distribution of the arc will be used in more detail later on;
 if we only fit the position of the arc we simply choose two of these
 corresponding points (A11 and A12) on opposite sides of the critical
 line, require that they and A2 have a common source (A) and that the
 parity of $A11$ and $A12$ is reversed. We used the positions of the
 knots (which have the same surface brightness) marked in Fig. 3 as
 $A11$ and $A12$.  The use of the $A11$ and $A12$-positions together
 with the parity constraint requires the critical line to pass between
 A11 and A12, e.g., it rejects models where $A11$ and $A12$ belong
 to the same large segment of a giant tangential arc. The parity
 constraint is easily implemented by adding a term to (14)
 proportional to the product of the determinant of the Jacobians if
 this product is positive.

Thus, generally four kinds of observables can be used for the model
fit: positions (A11, A12, A2, B1, B2, B3, W1, W2, WC, W3, C1, C2, C3),
flux ratios, the parity of the lens map at A11/A12 and B1/B2 and the
flux-distribution of the arc. In Table 2 one can read off all
parameters used for the lens modelling, and whether they are varied
($+$) or kept constant ($-$). In the first case the number below the
plus equals the best fit value, in the second case it equals the
assumed fixed value.  As discussed already, the lens model is
described by the cluster ($\sigma_{\rm cl}$, $\epsilon_{\rm cl}$,
$\phi_{\rm cl}$ and $\theta_{\rm cl}$), the cD galaxy ($\sigma_{\rm
cD}$, $\epsilon_{\rm cD}$, $\phi_{\rm cD}$), the possibly perturbing
galaxies DB, V and S next to B1/B2, A1 and left of W3 ($\sigma_{\rm
DB}$, $\sigma_{\rm V}$, $\sigma_{\rm S}$), and the unknown lensing
strengths
\begin {equation}
d_B:={D_{\rm ds}(B) \over D_{\rm s} (B)}\eck{ D_{\rm ds}(cB58)\over 
D_{\rm s}(cB58) }^{-1} \;,\;
\end{equation}
\begin{equation}
d_W:={D_{\rm ds}(W) \over D_{\rm s} (W)}\eck{ D_{\rm ds}(cB58)\over 
D_{\rm s}(cB58) }^{-1} \;,\;
\end {equation}
\begin{equation}
d_C:={D_{\rm ds}(C) \over D_{\rm s} (C)}\eck{ D_{\rm ds}(cB58)\over 
D_{\rm s}(cB58) }^{-1} 
\end {equation}
of the lens for B, W and C relative to cB58. Here $D_{\rm ds} (X)$ and
$D_{\rm s} (X)$ are the angular diameter distances from the cluster to
a source X and from the observer to the source.  The starting values
used in the minimization are given in the first line of Table 3. They
are motivated by prior knowledge in the case of $\sigma_{\rm cl}$,
$\sigma_{\rm cD}$ and $\sigma_{\rm DB}$: i.e. by the best fit value
for the velocity dispersion of Carlberg et al. (1996a,b), the
measurement of the velocity dispersion of the cD by Ziegler \& Bender
(1997), and the estimate of $\sigma_{\rm DB}$ using the
$D_n$-$\sigma$-relation which gives $84\pm15 {\rm km/s}$. If nothing
else is stated we restrict the allowed range for the velocity
dispersion of the cD in the minimization of (14) to $246 \; {\rm
km/s}\le \sigma_{\rm cD} \le 306 \; {\rm km/s}$ which is the
1-$\sigma$ interval for the observed uncorrected and
aperture-corrected velocity dispersion of the cD. The ellipticity of
the cluster and the cD is limited by $\epsilon_{\rm cl}\le 0.25$ and
$\epsilon _{\rm cD}\le 0.2$.  The role of the V-galaxy will be
discussed later; a velocity dispersion of $\sigma_{\rm V} \le 180 {\rm
km/s}$ will be a safe upper limit irrespective of its redshift. The
limiting values for $d_B$, $d_W$ and $d_C$, $0.1 \le d_X \le 1.1$ only
imply that the galaxies of the B, W and C-system are behind the
cluster and below a redshift of five. To illustrate that, we have
plotted the lensing strength for a galaxy at redshift $z$ in units of
the lensing strength for cB58 for three different cosmological models
in Fig. 9 (for details concerning filled beam angular diameter
distances see Fukugita et al. 1992, Asada 1996).  In terms of lensing
strength, a galaxy at $z=2.7$ is almost at `infinity' for a cluster at
redshift $z=0.37$; the lens is stronger only by $8\%$ for a galaxy at
redshift of five, nearly independent of the cosmological model
assumed.
\label{fig:.1}
\begin{figure}
\begin{center}
  \epsfxsize=\hsize\epsffile{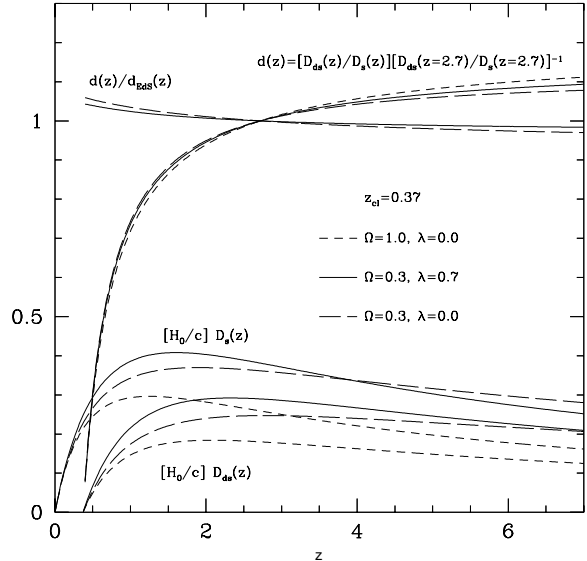}
\end{center}
\caption{For three different cosmological models we have plotted the
angular diameter distances from the observer (O) and the
angular diameter distances from the cluster (C) to
redshift $z$ in units of the Hubble length $(c/H_0)$: Einstein-de Sitter model
(short-dashed), a low-$\Omega$ universe with $\lambda=0$ (long-dashed)
and with $\lambda=1-\Omega$. The redshift of the cluster is $z_{\rm
cl} =0.37$.  The remaining three curves which start at redshift
$z=0.37$ and are equal to zero there show the lensing strength $D_{\rm
ds}/D_{\rm s}$ of a galaxy at redshift $z$ in units of the lensing
strength of the galaxy cB58 at $z=2.7$, for the same three
cosmological models. This ratio is called relative lensing strength
$d$ further on.  The two flat curves show the relative lensing
strength for the two low-$\Omega$ cases in units of the relative
lensing strength for an Einstein-de Sitter universe.
}
\end{figure}

Minimizing $E$ in (14) implies that the distances between the source
 positions -- estimated for all the members of a multiply lensed
 system -- are minimized in the source plane (In the ideal case the source
 separation is zero, because the galaxies are assumed to have a common
 source.) To estimate the quality of the best fitting model $\hat {
 \vec p}$ we define the `mean' source for each image system as a disk
 with radius of half a pixel centered on
\begin {equation}
\hat {\vec  \beta_k}
 := {1\over \sum_{i_k}^{I(k)} w(i_k)}\;  \sum_{i_k=1}^{I(k)} \vec \beta_{i_k}
( \hat {\vec p}) \; w(i_k) 
\quad .
\end{equation}
With ray tracing we approximately invert the lens equation and solve
for the extended images corresponding to the mean source. Let
$\vec \Delta(i_k)$ be the minimum distance of  $\vec \theta_{i_k}$ from
the inverted extended  images; we then define
\begin {equation}
\Delta^2_k:= \sum_{i_k=1}^{I(k)}   
\abs{\vec \Delta(i_k)}^2 \; w(i_k)
\end {equation}
as the quadratic error of the fit.  This error (divided by the number
of images used for the fit) has to be compared to the error of image
localisation or the extent of the images at $ {\vec \theta}_{i_k}
$. Actually $\Delta^2_k$ should be minimized with respect to $\vec p$
to obtain the best fitting model. Instead of this time consuming
process we minimize E (14) and assume that the parameters which
minimize $E$ are not too  different from those which minimize
$\Delta_k^2$.
The fit quality of a model is quantified in Table 3. The first four
columns contain the quadratic error of the fit in the lens plane which was
defined in equation (19). If a multiple {\it image system} (A, B, W or
C) is used for the fit, then the weights in (18) are equal to that
used in the minimization; that means, if one of the multiple images of
the system is not used for the fit, it is also not used to estimate the
fit quality. If an image system is not used for the fit, then
$\Delta^2$ measures the quality of the prediction that the observed
images have a common source; in this case (if nothing else is stated)
all of its multiple image positions enter Eq. (19).

\subsection{Results}
\label{sec:5.3}

The models explored followed a strategy of increasing complexity,
aiming at assessing the robustness of the lensing predictions.
The first models we investigated used the positions of A1/A2 and B1/B2
as observables;
good fits predict that C1/C2/C3 and W1/W2/WC are multiple images; the
exact position of W3 depends on the slope of the mass profile in the
center; since W3 has the same `tail' (in color and morphology) as
W1, W2 and WC with reversed parity relative to W1, it is obviously a
fourth image of WC, and it is used for further constraining the mass
distribution in the core of MS1512$+$36. The observations show some
evidence that the fifth image is merging `head on head' with the
fourth image (see Fig. 7).  If this is the case then the radial
critical line of the combined mass distribution (cluster and cD
galaxy) passes through the head of W3. Alternatively, a slightly
different redshift for A and W is sufficient to allow that an
additional image of A (the existence of which is predicted for some of
the models) lies next to the head of W3. Therefore we leave the fifth
image position of the W-system unconstrained. Since the galaxy B3 is
too faint for a precise measurement of its magnitude in R, 
and on the other hand is surprisingly bright relative to B1 and B2
(which are near a critical line and thus should be magnified
considerably) we can not be absolutely sure that B3 is indeed the
counterimage of B1 and B2.  Consequently, 
 we use as a first step only the positions
of A1/A2, W1/W2/WC/W3, and C1/C2/C3  in the models MIa\&b. In a second
step (model MIIa), we reverse the approach and check what happens when
the positions of A1/A2, W1/W2/WC/W3, and B1/B2/B3 are used. In models 
MIIIa and MIVa we explore the role of the slope of the
potential. Models MVa\&b\&c all the positions are used.
\label{fig:10}
\begin{figure}
\begin{center}
  \epsfxsize=\hsize\epsffile{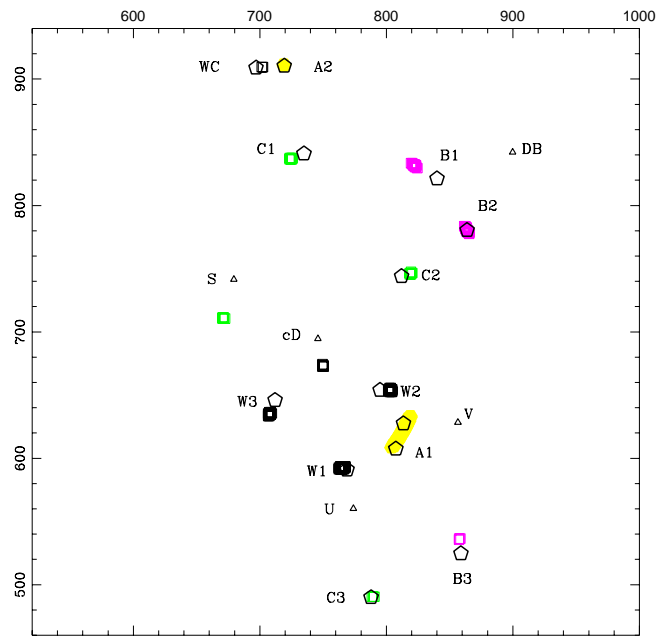}
\end{center}
\caption{Observed positions of multiple images (hexagons) versus
prediction (hatched regions) using the mean sources defined in (18).  
The best fitting
model MIa was obtained using only the positions of the image system A,
W and C and considering only the light deflection caused by the cD and
the cluster as `free' parameters. For more details, see text.  To
check the hypothesis that B3 is the counterimage of B1 and B2 we have
inserted the additional images predicted for B2 assuming its redshift
coincides with that of cB58.
}
\end{figure}

With model MIa we investigate whether it is sufficient to describe the
lens system with the deflection potential of the cD, the cluster and
DB, and whether models can be found where the potential depth of the
cD galaxy is in agreement with its observed velocity dispersion. Hence
the velocity dispersion of the cD is restricted to $246 \; {\rm km/s}
\le \sigma_{\rm cD} \le 306 \; {\rm km/s}$ (the $1 \sigma$-interval
for the uncorrected and aperture-corrected velocity dispersion). The
ellipticities of the cluster and cD potential are  assumed to be
smaller than 0.25 and 0.2, respectively; the velocity dispersion of DB
is kept constant at $\sigma_{\rm DB}=84 \; {\rm km/s}$. The resulting
best fit parameters are summarized in Table 2, and the predicted images
of the mean sources can be seen in Fig. 10. The arc is curved a bit
too strongly, but the position of the arc and its counterimage are fitted
perfectly. From Figure 10 and Table 3 we infer that the positions of
the predicted images of W (C) are off from the observed ones by 6.7
(6.9) pixels on average and thus the error is smaller than the typical
extent ($\ga 10$ pixels) of the images. To see whether B3 is indeed the
counterimage of B2 and B1 we mapped the image B2 back into the source
plane (assuming its redshift agrees with that of cB58) and we added
the predicted images of that source in Fig. 10 as well. Since the
positions of B1 and B2 were not used for the model fit one can not
expect that the predicted position of B1 agrees with that of the
observed one. The fact that the predicted third image of B is so close to
that of the `candidate' B3 adds further weight to the hypothesis that
B3 is the counterimage of B1 and B2.
\label{fig:11}
\begin{figure}
\begin{center}
  \epsfxsize=\hsize\epsffile{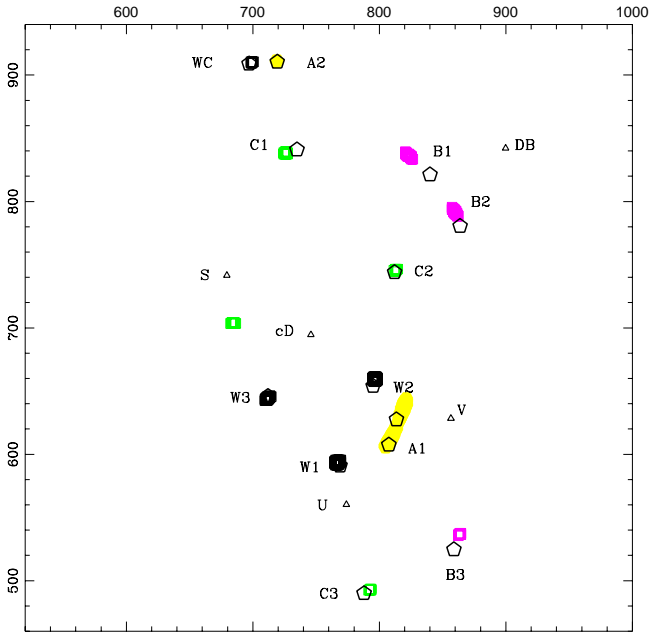}
\end{center}
\caption{For the model MIb we used the arc positions $\hat A11$ and
$\hat A12$ instead of A11 and A12. The remaining observables and the lens
modelling is the same as for MIa.  The arc becomes straighter (but
also too long) and the W-system fit improves. We calculated the mean
source of B1 and B2 and added the predicted images of that source .
As before, the prediction is off from the observation by more than the
extent of the galaxies B1-B3. However, it is obvious that B1 and B2
belong to a three image system, and that the third image is very close
to the observed B3.
}
\end{figure}

To check the robustness of the best fit parameters we used the
innermost two knots of cB58 (see Fig. 3, where they are marked as
$\hat A11$ and $\hat A12$) instead of A11 and A12 for the fit. The
resulting model is called MIb. The two knots together with the
reversed parity at $\hat A11$ and $\hat A12$ constrain the critical
curve at cB58 more strongly, because now the critical line must not
only be somewhere between A11 and A12 but more confined between $\hat
A11$ and $\hat A12$.  The quality of the fit improves: the fit for the
W and A system is almost perfect; the arc becomes straight due to the
increased ellipticity ($\epsilon_{\rm cD}=0.199$) of the cD
galaxy. The best fit parameters of MIa and MIb differ and they
demonstrate that an increase of the velocity dispersion from $\approx
570$ to $630$ km/s can be compensated by an increase of the core
radius from $120$ to $180$ pixels (which flattens the mass profile in
the center). As far as the arc position and direction are
concerned, a more circular cluster profile can be compensated by an
increase of the ellipticity of the cD galaxy's deflection potential.

We point out that  taking into account flux ratios of multiple
images can not resolve this ambiguity: objects near the critical line
(B1, B2, A1) can not be considered, because their flux ratio depends
on the details of the mass distribution near the critical line and
they cannot  be treated as point sources. Thus, only for the objects C1,
C2, C3 and W1, W2 and WC one can compare directly the magnification
ratios predicted by the model with the observed flux ratios. From
these objects, the magnitudes of W2 and C2 are affected by red light
and absorption in V due to the cD galaxy (or by inaccurate subtraction
of the cD light); since C1 and C3 differ in color by about $0.6$, only
W1 and WC remain as `clean' sources. In the coadded data (which
improve the signal-to-noise especially for WC) we obtain a magnitude
difference of $\Delta^W_{C1}=2.2\pm 0.75$ for the objects WC and W1. The
comparison of the predicted magnitude difference of the multiple
images $\Delta^C_{21}$, $\Delta^C_{31}$ and $\Delta^W_{C1}$ in Table 3
shows that the predicted flux {\it ratios} are similar and agree
reasonably well with the observations, where again MIb is a slightly better
fit. The magnification of A2 increases from $2.15$ to $2.89$ if one
changes the parameters from MIa to MIb, since then the larger velocity
dispersion provides a larger surface mass density (and shear) at A2.

Since the models MIa\&b predict the counterimage of B1 and B2 very
close to B3 we now use the same observables as for MIa and 
assume further that B1, B2 and B3 have a common source; 
the free parameters are as in MIb. 
The resulting best fit is denoted by MIIa. The quality of the
fit is good (see Table 3 and Fig. 12). The C-system is predicted very
accurately if we use the value for $d_C\approx 1.03 $ derived from
MIa\&b. Of course, a zero velocity dispersion for DB is
unrealistic. It is not unexpected, however, that the best fitting
value for the velocity dispersion is off from a reasonable one (say
$84{\rm km/s}$) because DB is relatively far away from the critical
line. For a velocity dispersion of $84\;{\rm km/s}$ the shear (and
surface mass density) from DB at B1/B2 is only about 0.02. One can not expect
that describing the true deflection potential by the sum of two
elliptical potentials provides an accurate fit not only for the
observables in the core but also at the positions of B1 and B2 at the
2 percent level, since the ellipticity and the slope of the surface
mass density is expected to change from the inner to the outer parts
of the cluster. We have tested how large the deviation from the model
potential must be to allow for a perfect fit for the positions of B1,
B2 and B3, by applying the same model fit as in the case of MIIa, but
now allowing for a `negative' mass at DB; a `negative' mass
corresponding to $-70$ km/s changes the shear and magnification
locally by only $0.04$ (relative to the $+84$ km/s-case) -- which is
unmeasurable by weak lensing methods -- and improves the fit quality
to $\Delta^2_B=9$ relative to MIIa and dims the brightness of B1
and B2 relative to B3.

\begin{figure}
\begin{center}
  \epsfxsize=\hsize\epsffile{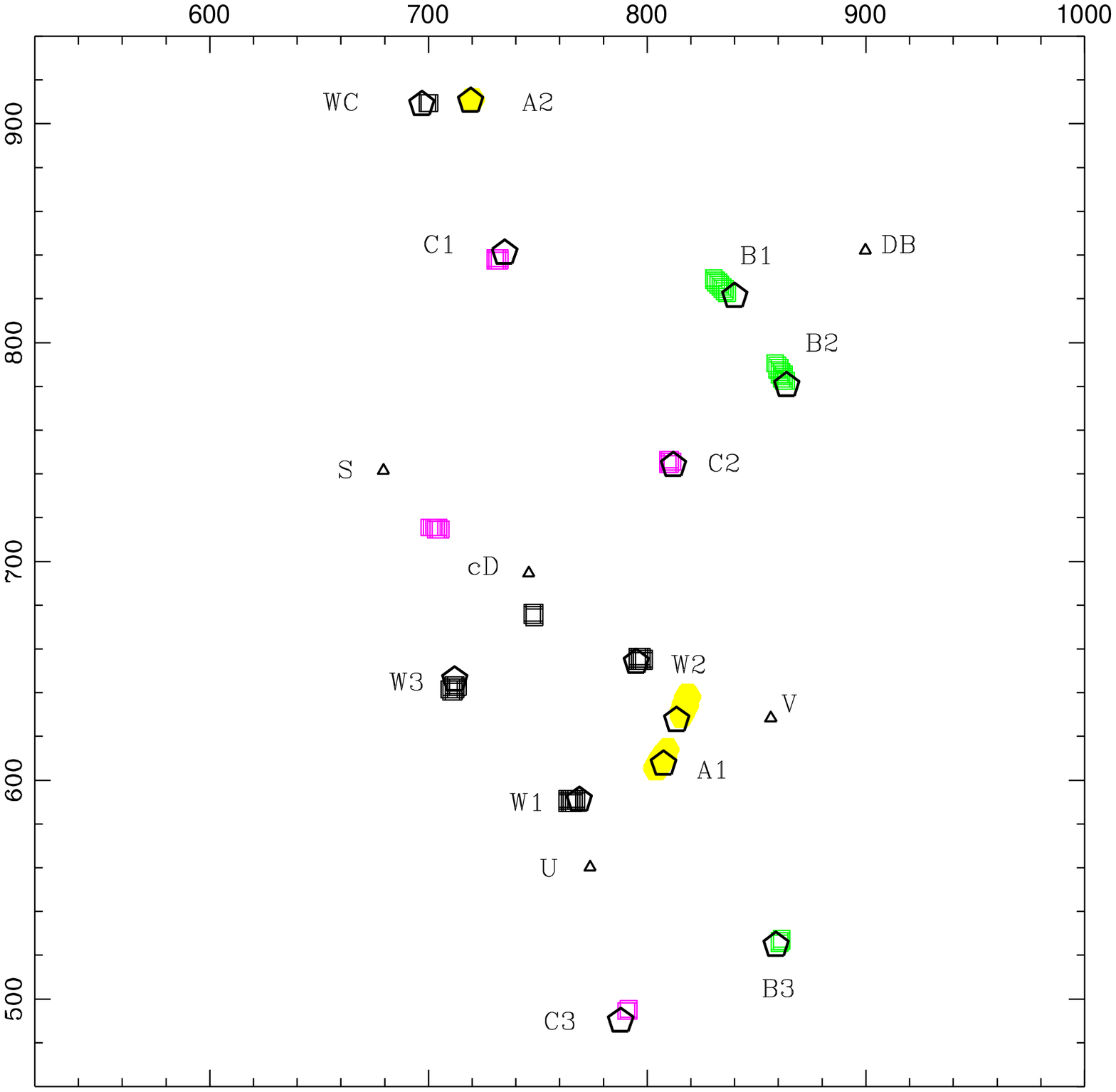}
\end{center}
\caption{For this fit the free parameters of the lens model were the
same as for MIa\&b with an additional variation of the velocity
dispersion of DB.  Instead of the C-system the B-system was used for
the fit. Assuming that $d_C=1.03$ the mean source of C1, C2 and C3 is
obtained and the predicted images of that source are also plotted.
}
\end{figure}
%
%
%
%
%            tabelle 2
% 
\begin{table*}
\caption{Best fit values ($+$) and assumed fixed values (--) for the
cluster, the galaxies and the redshift of the objects. Velocity
dispersions $\sigma_{\rm cl}$, $\sigma_{\rm cD}$, $\sigma_{\rm E}$ are
given in km/s. Major axis position angles $\Phi_{\rm cl}$ and $\Phi_{\rm
cD}$ are given in degrees. The core radii $\zeta_{\rm cl}$ are quoted
in  pixels -- where one pixel corresponds to $0\farcs 0498$. See
Fig. 9 to convert the lensing strengths $d_B$, $d_W$, $d_C$ into redshifts.}
\label{tab:4.1}
\begin{center}
\begin{tabular}{|l |l |l |l |l |l|l |l |l |l |l |l
 l| l| l|  }
\hline
Model & $\sigma_{\rm cl}$ & $\epsilon_{\rm cl}$ & $\phi_{\rm cl}$ & 
$\zeta_{\rm cl}$ & $\sigma_{\rm cD}$  &
$\sigma_{\rm DB}$ & $\sigma_{\rm S}$ & $\sigma_{\rm V }$ 
& $d_B$ & $d_W$ & $d_C$ &
 $\epsilon_{\rm cD}$ & $\phi_{\rm cD}$   & $\sigma_{\rm E}$
\\
\hline
\hline
start   & $710$ & $0.0$ & $8$ & $100$ & $286$  & $84$ &$0$ & $0$ 
& $1$ & $1$ & $1$ 
& $0.06$   & $10$  &$176.0$
\\
\hline
\hline
  & $+$ & $+$ & $+$ & $+$ & $+$ & -- &--&--
 & --&  $+$   &  $+$  & $+$  & $+$  & --
\\
MIa$^{(1a)}$     & $569.1$ & $0.1579$ & $11.34$ & $122.0$ & $246.3$  & $84$ &$0$  & $0$ & 
$1.$  
& $1.002$ & $1.048$  & $0.0816$ & 12.13 &$176.0$ 
\\
 MIb$^{(1b)}$     & $630.8$ & $0.1091$ & $6.465$ & $181.3$ & $251.4$  & $84$ &$0$  & $0$ & 
$1.$  
& $0.9998$ & $1.026$  & $0.1991$ & 26.7 &$176.0$ 
\\
\hline
  & $+$ & $+$ & $+$ & $+$ & $+$ & $+$ &--&--
 & $+$ &  $+$   &  --  & $+$  & $+$  & --
\\
 MIIa$^{(2)}$     & $564.0$ & $0.2023$ & $12.14$ & $141.0$ & $246.0$  & $0$ &$0$  & $0$ & 
$1.025$  
& $1.0$ & $1.03$  & $0.0001$ & 12.13 &$176.0$ 
\\
\hline
  & $+$ & $+$ & $+$ & $+$ & $+$ & $+$ &--&--
 & $+$&  $+$   &  $+$  & $+$  & $+$  & --
\\
 MIIIa$^{(3)}$     & $593.4$ & $0.1797$ & $9.73$ & $158.27$ & $249.1$  & $72$ &$0$  & $0$ & 
$1.020$  
& $1.0004$ & $1.037$  & $0.0$ & -- &$176.0$ 
\\
\hline
  & $+$ & $+$ & $+$ & $+$ & $+$ & -- &--&--
 & -- &  $+$   &  --  & $+$  & $+$  & --
\\
 MIVa$^{(4)}$     & $585.0$ & $0.1894$ & $12.19$ & $152.5$ & $247.0$  & $84$ &$0$  & $0$ & 
$1.020$  
& $1.005$ & $1.035$  & $0.0018$ & 18.54&$176.0$ 
\\
\hline
  & $+$ & $+$ & $+$ & $+$ & $+$ & $+$ &--&--
 & $+$ &  $+$   &  $+$  & --  & --  & --
\\
 MVa$^{(5)}$      & $571.7$ & $0.1753$ & $11.77$ & $144.9$ & $268.4$  & $70$ &$0$  & $0$ & 
$1.024$  
& $1.001$ & $1.042$  & $0.06$ & 10.00 &$176.0$ 
\\
 MVb$^{(5)}$     & $583.2$ & $0.1511$ & $12.40$ & $136.1$ & $246.0$  & $70$ &$0$  & $0$ & 
$1.024$  
& $1.002$ & $1.042$  & $0.11$ & 10.00 &$176.0$ 
\\
 MVc$^{(5)}$     & $615.4$ & $0.1407$ & $11.68$ & $166.6$ & $246.0$  & $70$ &$0$  & $0$ & 
$1.023$  
& $1.004$ & $1.035$  & $0.06$ & $\phi_{\rm cl}$ &$176.0$ 
\\
\hline
  & $+$ & $+$ & $+$ & $+$ & $+$ & $+$ &$+$ & $+$
 & $+$ &  $+$   &  $+$  & $+$  & $+$ & $-$
\\
% MVIa$^{(6)}$     & $549.5$ & $0.1980$ & $7.12$ & $142.7$ & $246.2$  & $18$ &$%144$  & $115$ & 
%$1.016$  
%& $1.002$ & $1.036$  & $0.1819$ & $47.64$ &$176.0$ 
%\\
% MVIb$^{(6)}$     & $548.2$ & $0.2071$ & $8.55$ & $139.0$ & $246.0$  & $0$ &$1%31$  & $116$ & 
%$1.014$  
%& $1.005$ & $1.038$  & $0.1391$ & $49.82$ &$176.0$ 
%\\  
%             das a model ist ein c model bei mir
%
 MVIa$^{(6)}$     & $582.0$ & $0.1581$ & $14.30$ & $159.9$ & $246.2$  & $7$ &$111$  & $104$ & 
$1.016$  
& $1.007$ & $1.035$  & $0.1301$ & $12.22$ &$176.0$ 
\\
\hline
\hline
\multicolumn{15}{l}{(1a) Positions of A11/A12/A2, W1/W2/W3/WC, 
C1/C2/C3 fitted.}\\
\multicolumn{15}{l}{(1b) Positions of $\hat A11/\hat A12/A2$,
W1/W2/W3/WC, C1/C2/C3 fitted.}\\
\multicolumn{15}{l}{(2) Positions of A11/A12/A2, W1/W2/W3/WC
, B1/B2/B3 fitted.}\\
\multicolumn{15}{l}{(3) As (1a), with triple weight for C1/C2/C3.}\\
\multicolumn{15}{l}{(4) Positions of A11/A12/A2, W1/W2/W3/WC fitted.}\\
\multicolumn{15}{l}{(5) Positions of A11/A12/A2, W1/W2/W3/WC, B1/B2/B3,
C1/C2/C3 fitted.}\\
\multicolumn{15}{l}{(6) Positions of A11/A2, W1/W2/W3/WC, B1/B2/B3,
C1/C2/C3 fitted as well as the shape of the arc cB58.}\\
\end{tabular}
\end{center}
\end{table*}
%
%
%
%
%  tabelle 3
%
\begin{table*}
\caption{The quality of the best-fit models. The quantities
$\Delta^2_W$, $\Delta^2_A$, $\Delta^2_C$, $\Delta^2_B$ give the square
position residuals (in pixels) as defined in Eq. 19; $\mu^A_2$  are
the predictions for the
magnification of A2 (counterimage of cB58); $\delta_{21}^C$,
$\delta_{31}^C$,
$\delta_{C1}^W$,  $\delta_{32}^B$ are the predicted magnitude
differences for C2/C1, C3/C1, WC/W1 and B3/B2. The first line contains
the observed magnitude differences in the V-band. The uncertainties
for  these
magnitude differences can be obtained using table 1.}
\label{tab:4.3}
\begin{center}
\begin{tabular}{ l| l| l| l| l| l| l|
 l| l| l| l| }
\hline 
Model & $\Delta ^2_W $ & $ \Delta^2_A$ & $\Delta^2_C$ & $\Delta^2_B$ & $\mu^A_{2}$ 
& $\delta^C_{21}$ & $\delta^C_{31}$ &$\delta^W_{C1}$ &$ \delta^B_{32}$
\\
\hline
%Observed & -- & -- & -- & -- & -- & 0.38 & 0.35 & 2.47 & 1.68 \\ 
Observed & -- & -- & -- & -- & -- & 0.33 & 0.92 & 2.26 & 1.68 \\ 
\hline
MIa  & 179.2
& 0.5 & 144.4 & --     & 2.153 & 0.196  & 0.674 &1.63 & 1.05
\\
MIb  & 20.6
& 4.0 & 79.51 & 396.8  & 2.894 & 0.337  &  0.775 &1.79 & 0.799
\\
\hline
MIIa  & 22.7
& 6.0 & 40.8 & 20.8  & 2.180 & 0.21  &  0.698 &1.79 & 1.27
\\
\hline
MIIIa  & 226.4
& 54.0 & 11.8 & 1.2  & 2.374 & 0.19  &  0.734 &1.68 & 1.42
\\
\hline
MIVa  & 41.3
& 1.0 & 118.5 & 352.7   & 2.281 & 0.17  &  0.693 &1.72 & 1.2
\\
\hline
MVa  & 74.7
& 5.0 & 62.0 & 190.1   & 2.291  & 0.23   &  0.673 &1.64 & 1.17
\\
MVb  & 46.2
& 10.6 & 101.3 & 269.3   & 2.323  & 0.16   &  0.681 &1.66 & 1.11
\\
MVc  & 85.3
& 0.04 & 54.0 & 278.1   & 2.766  & 0.20   &  0.79 &1.92 & 1.19
\\
\hline
%\hline
%MVIa  & 7.2
%& 1.2 & 147.1& 281.6  & 2.250 & 0.27   &  0.72 &1.56 & 1.23
%\\
%MVIb  & 24.8
%& 13.2 & 75.9 & 61.9  & 2.121 & 0.20   &  0.65 &1.51 & 1.35
%\\
%           das a model ist mein c model
%
MVIa  & 43.6
& 1.4 & 116.4 & 246.0  & 2.386 & 0.12   &  0.71 &1.63 & 1.28
\\
\hline
\hline
\end{tabular}
\end{center}
\end{table*}

When the positions of the As, Ws and Cs are used for the model fit,
the prediction for B is not as good as when the positions of the As,
Ws and Bs are used to predict the Cs. The reason is that in the first
case the mass profile is probed mainly in the central region, whereas
in the latter case the larger distances of B1/B2/B3 allow to constrain
the slope of the mass profile better.  We demonstrate this with the
model MIIIa, which was derived in two steps: Best fit values for the
parameters describing the cD galaxy, the cluster and the redshifts
where obtained as for MIb, but with a triple weight for the positions
of C1, C2 and C3. Using these parameters, the velocity dispersion of
DB and redshift of B where varied to optimize the predicted positions
for the Bs as well. The best fitting values are $d_B=1.02$ and
$\sigma_{\rm DB}=72 \;{\rm kms/s}$. As expected (see Table 3) the
quality of fit improves in the outer regions and decreases in the
core.

Thus we have demonstrated that the numerous multiple images can be
fitted at the same time with a very simple model consisting of a
singular (non-singular) elliptical potential for the cD galaxy (and
the cluster). The fact that the best fit parameters in Table 2 can be
quite different from model to model shows that no fine tuning of the
parameters is necessary to explain the multiple images. (However,
model MIa\&b is not a good fit for the image positions of the
B-system). The images C are most robust with respect to a change of
the parameters, whereas the W and A galaxies are sensitive to the
steepness and height of the combined potential (of course the
contributions coming from the cD and the cluster can be varied), and
the Bs are more sensitive to the ellipticity and slope of the
potential at $\theta \ge 10\arcsec$.

\subsection{The magnification of A2 and limits to the cluster's velocity
dispersion and core radius}
\label{sec:5.4}
Since not necessarily the whole source of cB58 has to be lensed into the
gravitational arc cB58, and the magnification near critical lines is very
model dependent, the magnitude of the unlensed source of cB58 can be
obtained most accurately from the counterimage A2.

The most pessimistic statement concerning the magnification of A2 is
obtained by considering the so-called mass-sheet degeneracy
(e.g. Schneider \& Seitz 1995) which states that all dimensionless
observables (source separation and flux ratios of multiple images) are
invariant under a transformation of the deflection potential according
to
\begin{equation}
\psi_{\rm tot}(x,y) \to \lambda \psi_{\rm tot}(x,y)
+(1-\lambda)(x^2+y^2)/2
\; ;
\end{equation}
in other words, a potential can not be disentangled from a rescaled
deflection potential (and thus a rescaled surface mass density and
shear) and the addition of a constant mass sheet with
$\kappa_0=1-\lambda$; the constant $\lambda$ is limited by the fact
the the total surface mass density must not be negative.

From the ground based data of Gioia \& Luppino (1994) we infer that
the cluster is optically poor and confined to a small region; there is
no hint that it is embedded in a larger scale structure which could
provide the unconstrained mass sheet. Thus the ignorance about the
mass-sheet degeneracy will add only a small uncertainty for the
magnification of A2 relative to the unlensed source.

The unconstrained core radius of the cluster, however, can mimic the
mass sheet degeneracy locally: increasing the velocity dispersion and
the core radius of the cluster is similar to rescaling the potential
and adding a constant surface mass density, as long as only a few
positions (that of the multiple images) are considered and these
positions are not spread enough in radius. If only the A and W system
are taken into account for the model fit, a large cluster velocity
dispersion can not be excluded, since then one can increase the core
radius to suppress the central mass peak; for a velocity dispersion of
$\sigma_{\rm cl}=840 \; {\rm km/s}$ one obtains a best fit core radius
of about $20\arcsec $ and $\sigma_{\rm cD}=278 \; {\rm km/s}$ with a
fit quality of $\Delta^2_W\approx 300$ and $\Delta^2_A=2$. The
magnification of A2 is then approximately 7. Even if one ignores the B
and C system, a model demanding such a large core radius seems to be
unplausible. On the other hand, if one uses only the positions A11,
A12, A2 and W1, W2, WC and W3, the limits for $\sigma_{\rm cl}$,
$\epsilon_{\rm cl,cD}$ as in the models before, and does not restrict
the velocity dispersion of the cluster, a low value of its velocity
dispersion and core radius is favored (the corresponding best fit
values are summarized as MIVa in Table 2). The magnification of A2
agrees with that obtained for earlier models in Table 3.

\begin{figure}
\begin{center}
  \epsfxsize=\hsize\epsffile{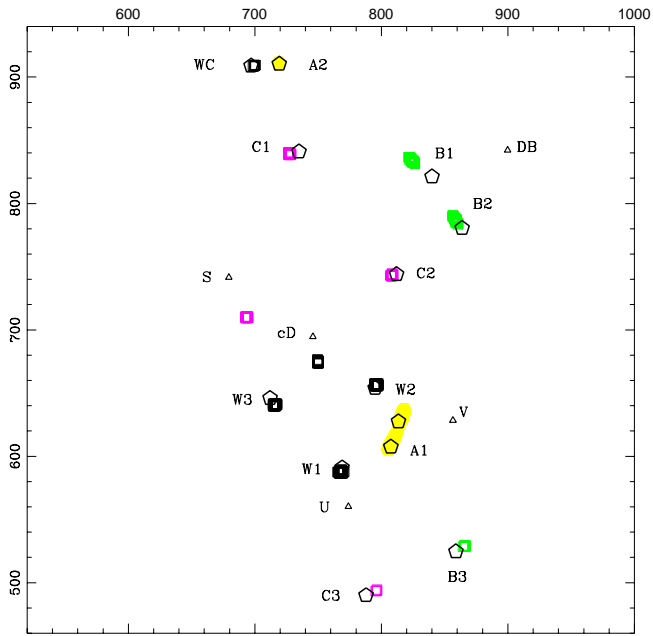}
\end{center}
\caption{For model MIVa we used only the positions of the A and
W-system to constrain the parameters describing the cD-galaxy and the
cluster. The allowed range of parameters was limited by $\epsilon_{\rm
cD}\le 0.2$, $\epsilon_{\rm cl}\le 0.25$ and 
$246 \; {\rm km/s}\;\le  \sigma_{\rm cl}\le 306 \; {\rm km/s}$. The
cluster velocity dispersion was not limited. The multiple image
positions for the C and B-system are predicted under the assumption 
$\sigma_{\rm DB}=84\;{\rm km/s}$, $d_B=1.02$ and $d_C=1.035$.}
\end{figure}

Including the positions of the B-system must constrain the maximum
velocity dispersion strongest. For a conservative estimate we ignore
that B3 is most likely the third image to B1 and B2. In this case, an
increase of the lensing strength for B1 and B2 for large velocity
dispersions can be avoided by shifting them to lower redshift. Without
a lower limit for the velocity dispersion of the cD, i.e.  allowing
for a redistribution of mass from the cD to the cluster and with
restrictions for the remaining parameters as for model MIIa, we obtain
a maximal acceptable velocity dispersion of the cluster of
$\sigma_{\rm cl}=670 {\rm km/s}$, with a core radius of $\approx
11\farcs 2$ and a cD velocity dispersion of $250.8$ km/s.  The
relative lensing strength of B then has to be $d_B\approx 0.9$,
corresponding to a redshift of 1.7 in an Einstein-de Sitter
universe. For this maximal velocity dispersion the magnification of A2
is equal to $\mu_{A2}=3.4$. If B3 is the counterimage of B1 and B2,
the maximal acceptable velocity dispersion has to be as low as
$600-610 {\rm km/s}$ to obtain a marginal fit with $\Delta^2_B\le
550$. Note that MIb can not be considered as a good fit for B1, B2 and
B3 and that in this case the sum in Table 3 includes only the
positions of B1 and B2. For a cluster velocity dispersion of $600 \;
{\rm km/s}$ the velocity dispersion of the cD and the core radius
become $257.8 \; {\rm km/s}$ and $8\farcs 8$. The magnification of A2
equals $2.56$ in this case.

These investigations show that in fact the magnification of A2 is
limited by $2.0\la \mu_{A2}  \la 3.4$ if B1, B2 are considered and
even more ($2.0\la \mu_{A2}  \la 2.9$\footnote{These limits were not
derived from a rigorous investigation, but contain all magnifications
for A2 predicted by models which include B3 in their fit.}
) if B3 is the third image of B1 and B2. 
Concerning the possible velocity dispersions of the cluster, 
the best fit parameters for the 
models MIa and MIb are almost as different as they can be to allow a
good fit. We have tested this by giving different weights to different
multiple image systems, by considering B1/B2/B3 instead of C1/C2/C3,
by demanding that the ellipticity and orientation of the cD-potential
equals that inferred from the light, by taking into account the
galaxies V and S as possible deflectors etc. From
this we infer limits of $540 \;{\rm km/s} \le \sigma_{\rm cl}\le
670\; {\rm km/s}$, $5\farcs 5 
\le \theta_{\rm cl}\le 11\farcs $ and $6 \degr \le
\phi_{\rm cl}\le 14 \degr$. Of course, these limits weaken, if one
uses only the positions of the A or W system for the fit.

\subsection{Alignment of  cD-light, cD-potential and the cluster potential}
\label{sec:5.5}
In the models MI-IV the ellipticity and orientation of the cD
potential were treated as free parameters. The best values for the
ellipticity and orientation for those models are spread between
$0-0.2$ and $\approx 11\degr - 19 \degr$. This does not imply that the
ellipticity and major axis of the cD potential is inconsistent with
the values inferred from the light using Fig. 2 and the relation $\hat
\epsilon_{\rm cl} = 3 \epsilon_{\rm cl}$.  Since lensing is only
sensitive to the combined potential of the cD and the cluster, the
orientation and ellipticity of the {\it individual} potentials are
undetermined within some range (see Table 2 ). To demonstrate the
consistency of the cD potential with the light distribution of the cD
we kept the cD parameters $\epsilon_{\rm cD}=0.06$ and $\phi_{\rm
cD}=10\degr$ fixed, varied those describing the cluster, the velocity
dispersion of the cD ($246\; {\rm km/s} \le \sigma_{\rm cD}\le 306 \;
{\rm km/s}$) and of DB ($70 \;\le \sigma_{\rm DB}\le 100\; {\rm
km/s}$), and the redshifts of the W, C and B system. We used the
positions of the A, W, B and C-system for the fit. With the exception
of the B-positions the best fit model (MVa) recovers the multiple
image positions very well. When $\epsilon_{\rm cD}=0.11$ is assumed
(model MVb) the fit quality reduces. For MVa\&b the major axes of the
potentials of the cD and the cluster are almost parallel. As a
consequence, the fit quality stays comparable to the models MVa\&b, if
one requires that the cD potential is strictly parallel to that of the
cluster: we obtain $\phi_{\rm cD}\equiv \phi_{\rm cl} \approx 12
\degr$ (see model MVc in Table 2). For the models MVa\&b the weights
$w(i_k)$ in equation (14) were equal for all positions used for the
fit. Since the multiple image positions are concentrated in the
cluster core (W1,W2,W3,A11,A12) the mass profile is constrained most
strongly there, and these observables are reproduced with the largest
accuracy. Giving a weight of 2-3 to the positions of the C-system
(B-system) and fitting the parameters in analogy to MVc, the accuracy
for the prediction of the C {\it and} B (B {\it and} C) system
increases.  This adds further weight to the hypothesis that both the
C- and B-system are triple image systems.  With the large weight to
the outer multiple images, the orientation of the major axis of the
cluster and cD potential decreases to 10 and 9 degrees.  This tilt of
the major axis of the potential is also seen in the light of the halo
of the cD galaxy: for $\theta \le 5\arcsec$ (A and W system) the
position angle equals about $10\degr$ and drops to $\le 8\degr$ in
that region where the images C and B occur ($\theta\le 7\farcs 5$).

We conclude, that the cD and the cluster are aligned and have a major
axis position angle of about $10\degr$, that the ellipticity of the
combined potential increases and the position angle of the major axis 
changes in the  outer parts of the cluster. The orientation of
the major axis of the optical light and the dark mass  thus
agrees with that of the X-ray light found by Hamana et al. (1997).

\section{The magnification of the arc}
\label{sec:6}
Up to now only two spots (A11 and A12) of the arc were used for the
lens modelling, and they were treated as positions of multiple
images. To make more use of the two-dimensional light distribution of
the arc for the lens modelling, we consider only (CCD)-pixels where
the counts exceed a rate of 5.5 and subdivide each pixel into two
triangles (see Schneider, Ehlers \& Falco 1992, page 300). 
For each model $\vec p$, all triangles of the arc are mapped
into the source plane. Using a $\approx 50$ times denser grid than in
the lens plane, we identify for every triangle in the image plane all
grid points in the source plane which lie within the corresponding
triangle in the source plane.  For a perfect model, each grid point in
the source plane is contained either in no or in two triangles. In the
second case one of these triangles has positive parity with respect to
its corresponding triangle in the image plane (which is outside the
critical line), and the other is mapped with negative parity onto the
image plane (and the corresponding triangle is inside the critical
line). Within the accuracy given by the finite spatial resolution of
the CCD pixels, the surface brightness of such triangles should also
agree. Gridpoints in the source plane which are singly imaged only
reflect an imperfect model (and observational errors). To obtain a
model which predicts the correct shape of the arc, we thus add a term
proportional to
\begin{equation}
\eta:= {N_s-N_d\over N_s+N_d}
\quad 
\end{equation}
to (14), where $N_s$ and $N_d$ are the number of gridpoints in the
source plane which are singly and doubly imaged, respectively. 
\label{fig:?.1}
\begin{figure}
\begin{center}
  \epsfxsize=\hsize\epsffile{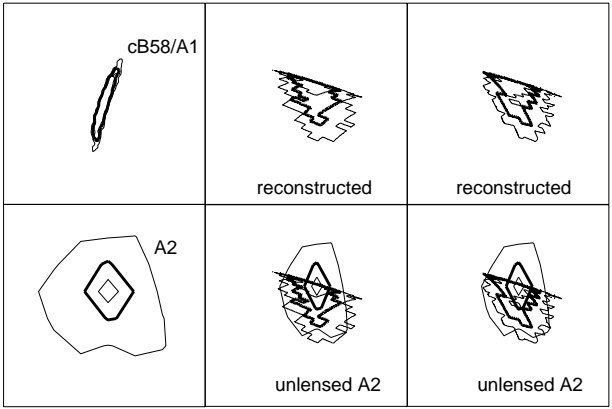}
\end{center}
\caption{In all six panels we plot surface brightness contours of  5.5, 10
(thick contour) and 15 counts per pixel. The upper left panel has a
side length of $5\arcsec$ and shows the surface brightness
distribution of cB58.  The remaining squares are $0\farcs 35$ on a
side. The upper middle and upper right panels show the 
reconstructed source (model MVIa) using the negative parity part  and
positive parity part  of the arc, respectively.
 In the lower left panel the contours
of the counterimage A2 can be seen. Using the surface density and shear
at the position of A2 we `unlens' the shape of A2 and obtain the
squeezed contours in the lower middle and right panels. On top of that
the corresponding contours of the reconstructed source of the arc are
overlaid.}
\end{figure}
Note
that adding a term proportional to $N_s$ or $-N_d$ instead of (21)
would introduce a bias towards high and low magnification of the
arc. The value of (21) is independent of the magnification of the arc
and equals (+1) and (-1) for a bad and for an optimal model. The best
fit parameters are obtained in two steps: first the positions of the
multiple images are used for the model fit and then the arc-shape
constraint (21) is added in (14) and the minimization is
continued. Whereas we generally use three positions (A11,A12,A2) to
fix the arc and its counterimage, we consider only A11/A2 or A12/A2
when the shape is taken into account. This guarantees on the one hand
that the global properties of the lens are changed only slightly and
in particular that the counterimage stays at the same position; on the
other hand, this allows changes in the mass distribution to turn
the arc into the correct direction.
\label{fig:??.1}
\begin{figure}
\begin{center}
  \epsfxsize=\hsize\epsffile{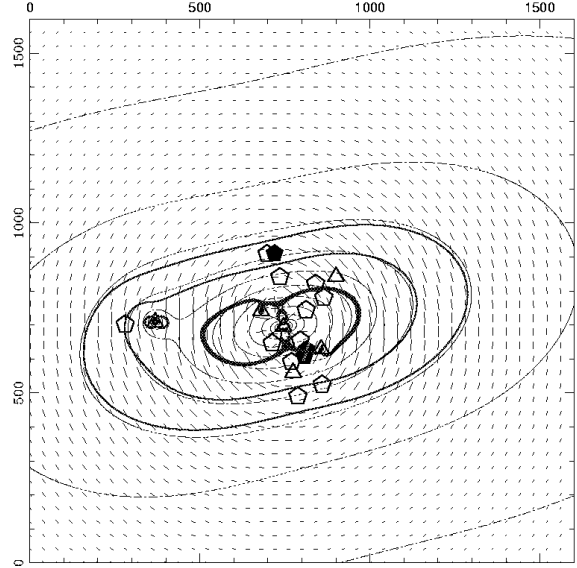}
\end{center}
\caption{Shear field (vectors), surface density contours (dotted
lines)  and magnification contours (solid lines)
for sources at redshift $2.7$ according to the
model MVIa. The surface density contours start at $\kappa=0.1$ and
are spaced
by 0.1. Filled (empty) pentagons denote the positions of A11/A12/A2
(and the remaining multiply imaged galaxies). Other galaxies like the
cD, the spiral, the elliptical E and the pertubers DB and V are added
as empty triangles. The three thick contours are that of magnification
2, 3 and `infinity', and thus the central thick contours denote the
critical lines for a galaxy at redshift $2.7$.
}
\end{figure}
In section 5 we showed that a good fit for the multiple image systems
can be obtained by combining the deflection potential of the cluster
and the cD galaxy. This is not the case for the arc shape: we can not
obtain a good fit for the arc shape {\it and} the position of the W
and A-system at the same time, because the arc demands a lot of shear
parallel to its major axis. A possible reason for the insufficient fit
can be inferred from Fig.1\&3: the arc is not entirely straight but
curved, and the center of curvature points into the opposite direction
that one expects if it is caused by the cD and the cluster. A possible
cause of this curvature is a perturbation of the light deflection by
the galaxy V on the right side of cB58: The galaxy V has the same blue
color but a much smaller surface brightness than cB58. Thus it could
be a blue, low surface brightness galaxy at low redshift or a `normal'
field galaxy at $z>1$.  We consider the first alternative as unlikely
due to the small number density of those galaxies.  If the redshift of
V is $\ge 2$, then it lies close to, or on the critical curve of the
lens corresponding to the redshift of V.  The `close' case can be
ruled out because there is no `second' image on the opposite side of
the critical line with comparable brightness. The fact that V has a
double core on the opposite sides of the $z\approx 2.7$ critical curve
would favor the `on the critical line' case. However, in that case a
much more elongated object similar to cB58 would be expected and a
counterimage on the right side A2 should be observable. The object 3.4
arcsec to the right of A2 is too far away and has too much of a red
color to be an option. We conclude that the galaxy V is at redshift
larger than one, but small enough to avoid being multiply imaged. In
this case a two lens-screen situation arises, where the light of cB58
is distorted by the galaxy V before it is deflected by the
cluster. Since we do not know the redshift of V we treat the light
deflection as if it takes place at the redshift of the cluster and
model V as a singular isothermal sphere. The best fit velocity
dispersion $\sigma_V$ can then not easily be related to the depth of
its potential, but describes only the `effective' lensing strength of
this galaxy. A two-screen model will only be useful once the redshift
of V is known. Most generally, the velocity dispersion $\sigma_V$ can
be considered as a one-parameter correction for the effective shear
and surface mass density at the arc.

For completeness we now also take into account the light deflection
caused by the spiral S. This is done to show that the additional light
deflection of the spiral does not change our preceding conclusions,
i.e. that the multiple images positions are in agreement with the
expectations from optical and X-ray light, and the dynamical
properties of the cluster galaxies.  Since none of the multiple images
is close to the spiral, it adds only a small perturbation to their
total deflection. Irrespective of its exact redshift ($z_{\rm S}
\approx 0.3$) we can therefore approximately describe its light
deflection as taking place at the redshift of the cluster. We
assume that the depth of its potential does not exceed $ 180{\rm
km/s}$, which corresponds to the 3-dimensional velocity dispersion of
a $1.8 L_*$-galaxy.

Fig.  14 shows the reconstructed sources for the arc cB58. It was
obtained using the arc shape (as observed), and the A, B, C and W
system with a low weight for the B-positions. The value of $\eta$
defined in (21) equals $-0.62$, and the remaining numbers
characterizing the fit quality for the positions are shown in table 3
(model MVIa). The area of the arc in the source plane equals $3.61 -
4.46$ (dithered) pixels, where the range is obtained from either
considering only the doubly imaged source pixels or all pixels in the
source plane.  This size is in agreement with that of the counterimage
A2 (16 pixels) if one takes into account that A2 is magnified by a
factor of $2.4$ and if about half of the total source is lensed into
the gravitational arc cB58 (as is indicated in Fig. 14 as well). From
the size ratios of the arc source ($3.61- 4.46$ pixels) and the
non-deconvolved arc (213 pixels) we obtain a magnification factor of
$\mu_{\rm arc} \approx 47.8-59.0$ within the contour of $5.5$ counts
per pixel. In Fig. 14 we have also unlensed the shape of A2 using the
predicted surface density and shear of the model MVIa at the position
A2. On the surface brightness contours of the unlensed object A2 we
have overlaid those of the reconstructed source of cB58. The contours
of $5.5$ counts per pixel agrees very well, although only the shape
and not the flux distribution of the arc was used for the model
fit. The contour at 10 counts per pixel is a bit larger for the
reconstructed arc source than for the unlensed counterimage. These
differences can be attributed to the finite pixel size in the image of
A2.  Fitting the shape of the arc adds only a local constraint to the
model and does not break the velocity dispersion-core radius
degeneracy discussed before. Thus the total magnification of the arc
will also increase if the velocity dispersion and core radius are
increased. The corresponding magnification of the arc can however be
estimated from the magnification of the counterimage: using that the
sizes of the arc and A2 are 213 and 16 pixels within the $5.5$ counts
per pixel contour, and that only a fraction $f\approx 1/2$ of the
total source is imaged into the arc cB58 one can roughly estimate the
arc magnification to be $\mu {\rm arc} \approx 213/16 \; \mu_{\rm
A2}/f \approx 25 \; \mu_{\rm A2}$.

In Fig. 15 the gravitational shear field, the surface mass density, and
magnification contours are plotted for the model MVIa. The shear field
is a measure for the direction and ellipticity that a circular source 
obtains through the lens effect of the combined mass distribution at the
same position. One can see that the shear field  left to
the elliptical E  is parallel to the $y$-direction of the chip, and
thus explains the large distortion of the arclet left of E (its
magnification equals $3$ for a redshift similar to cB58).
The surface density contours start at $\kappa =0.1$
and are separated by $\Delta \kappa =0.1$. Independent of the details
of the model the mean surface density in the chip-3 is $\ave{\kappa}
\approx 0.2$. The thick contours are that of magnification 2, 3 and
infinity (critical line for a source at $z=2.7$).

\section{Weak Lensing Analysis}
\label{sec:7}
We next consider the distortion of images of (presumably background)
galaxies near the core of MS1512.  With this analysis we do not aim at
a weak lensing mass reconstruction as in the case of Cl0939+4713
(Seitz et al. 1996) -- where the observations were deeper and the lens
stronger -- but at a consistency check of the weak lensing signal with
the cluster parameters estimated from multiple images. We use the
coadded V and R-data and consider only galaxies in the WFPC-fields
which are within a distance of 80 to 850 pixels (i.e. within a
distance of $42\farcs 5$) from the center of light of the cD
galaxy. The outer radius equals the maximum radius of a circular disc
centered on the cD galaxy which is completely within the data region.
A size cut of $50 \;$ (dithered) pixels is applied, and stars and
objects with a SExtractor-{\tt flag} larger than 16 are excluded. This
excludes objects with incomplete or corrupted aperture or isophotal
data.

The minimum flux of galaxies considered for the weak lensing analysis
was set equal to that of the completeness limit of the numbercounts in
the V+R-data. The bright cutoff was set 2.5 times this limiting flux.
Comparing the number counts for these galaxies in the WFPC2 field with
published results on the I-band counts (Smail et al. 1995), we
estimate that the selected galaxies have I-band magnitudes of about
$23.3 \la m_I \la 24.3$. According to the evaluation of the CFR-Survey
(Lilly et al. 1995) the mean redshift of those galaxies is predicted
to be of order one.

The shapes of the galaxies are estimated with the SExtractor software
package (Bertin \& Arnouts 1996). The width $\sigma_\chi$ of the
ellipticity distribution of the remaining 33 (29 of them are within
chip-3) galaxies is estimated and it is assumed (and checked) that
this width is basically unaffected by lensing. The deflection
potential of the cluster is described by (1), with the ellipticity,
orientation and velocity dispersion as free parameters, and the core
radius is assumed to be equal to $150\; {\rm pixel} $.  Let $\chi_i $
be the complex ellipticity as defined eg. in Schneider \& Seitz (1995)
of the $i$-th object, and $\ave{\chi}_i(\vec p)$ the model dependent
expectation value at the same position; further let $\sigma_\chi
t_i(\vec p)$ (for details on $t_i(\vec p)$ see also Schneider \& Seitz
1995) be the width of the ellipticity distribution expected at the
position of the $i$-th object for lensing parameters $\vec p$; we then
minimize:
\newpage
\begin{eqnarray} 
F
:=
 &-& \sum_{i=1}^{n_{\rm gal}} \ln  \Bigg\{ (1-f_{\rm cl}) \;
{1\over  t_i^2(\vec p)}
\exp\eck{-\abs{\chi_i - \ave{\chi}_i(\vec p) }^2\over \sigma_\chi^2
t_i^2(\vec p)}
\nonumber \\
&&\quad \qquad \qquad + 
 f_{\rm cl} \; 
\exp\eck{-{\abs{\chi_i }^2\over
\sigma_\chi^2}} \Bigg\}
\nonumber \\
\end{eqnarray}
\label{fig:6.1}
\begin{figure}
\begin{center}
  \epsfxsize=\hsize\epsffile{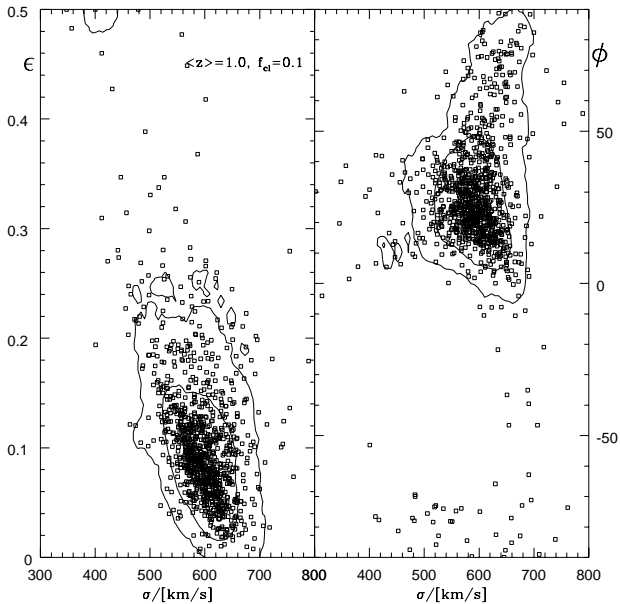}
\end{center}
\caption{Best fit estimates for the clusters velocity dispersion,
ellipticity and major axis position angle using 1000 bootstrapped
realizations of galaxies with $24 \le m_{V+R}\le 25$ within 850 pixels
distance to the cD galaxy. The contours approximately indicate that
region in which 68 and 90 percent of the best-fit values are
contained. The mean redshift of the galaxies and the contamination
with cluster members were assumed to be $\ave{z}\approx 1.0$ and $f_{\rm
cl}=0.1$ }
\end{figure}
with respect to $\vec p$; for the contamination of the galaxy sample
with cluster members (or foreground galaxies) we assume $f_{\rm cl}
=0.1$. To parametrize the distance to the galaxies we assume in
analogy to Eqs. (15)-(17) a relative lensing strength of $d=0.72$
which corresponds to a mean redshift of $\ave{z}\approx 1$ in an
Einstein-de Sitter universe. For 1000 bootstrapping realizations of
the data catalog we determine the best fitting parameters for the
velocity dispersion, the major axis position angle and the ellipticity
of the clusters deflection potential. The result is shown in Fig. 16
where each point in the $\epsilon_{\rm cl}$-$\sigma_{\rm cl}$ and
$\phi_{\rm cl}$-$\sigma_{\rm cl}$ plane corresponds to a best fit for
one bootstrapped data catalog. We added contours in which
approximately 90 and 68 percent of the best fit parameters are
contained. According to Fig. 16 the 1-$\sigma$ interval for the
velocity dispersion and the major axis position angle equals
$\eck{520,670}\; {\rm km/s}$ and $\eck{5\degr,50\degr}$, with best fit
values of about $600 \; {\rm km/s}$ and $20\degr$.  We point out that
the estimate for the velocity dispersion depends on the mean redshift
assumed for the galaxies and on $f_{\rm cl}$.  If that former was
equal to 0.8 instead of 1, the estimated velocities increase by 10
percent. Another uncertainty is the unknown core radius since the data
field ($42\arcsec$ radius) is not very large compared with the core
radius ($150\; {\rm pix} =7\farcs 5$). If the core radius is allowed
to vary between 120 and 200 pixels, the distribution of the points the
$\epsilon$-$\sigma_{\rm cl}$-plane becomes broader, in particular
towards larger velocity dispersions, since a larger core radius can be
compensated by an increased velocity dispersion. But even in this case
both the 68 and 90 percent upper confidence limits for the velocity
dispersion are still below $800 {\rm km/s}$. The assumption of a core
radius of 150 pixels is, by the way, not unreasonable because it 
agrees with that found from the the X-ray observations of this cluster
(Hamana et al. 1997).

\section{Summary and Discussion}
We have shown that the `protogalaxy'  cB58 owes its large apparent
brightness to the lens effect caused by the cluster
MS1512 and  its cD galaxy. The symmetry in the surface brightness
distribution reveals that cB58 is a gravitational fold arc; the large
axis ratio is caused by the stretching of the source parallel to the
major axis of the arc.

Three systems of multiply imaged galaxies are found. In the first case
(W), a shrimp-like object is mapped into five images, of which four can be
identified. The redshift of the W-system must be very similar to that
of cB58. Measuring the flux ratios and colors of the galaxies of the
W-system is difficult, because all objects are small, not much above
the noise level, or their colors are affected by red light and
absorption caused by the cD galaxy.

The second system (C) consists of three images  which also have not
well determined and slightly inconsistent colors. The three images
of the system have the same  head-tail morphology, their positions can
most robustly be fitted  for different parameters of the lens models, 
and the relative inclinations of their head-tail axes agrees with the 
predictions from lens models. For most models the source redshift is
$z\approx 3.5$. 

The third system (B) is also a triple system; two images  are close to
the tangential critical line and their magnification is affected by a nearby
elliptical galaxy (DB) with an estimated velocity dispersion of
$\sigma_{\rm DB}=84 {\rm  km/s}$. The estimated redshift of the source
is $\approx 3$. However, a precise prediction for the redshifts of the
B and C-system is difficult, since the
lensing strength increases only
slowly with redshift (see Fig. 9) and thus a broad range of redshifts
is formally possible.  An upper limit of $z<4$ for the B and C system
is set by their detectability in the B-band observations of  Gioia \&
Luppino (1994).

The lens system was modelled with a singular (non-singular),
elliptical isothermal potential for the cD galaxy (for the cluster)
and isothermal potentials for additional galaxies. Because the
multiple images are not spread much in radius, a core radius-velocity
dispersion degeneracy arises; if only the inner image systems (W and
A) are considered, a cluster velocity dispersion of 840 km/s
can not be excluded if a core radius as large as $\approx 20 \arcsec$
is accepted. Without any restriction to the cluster velocity
dispersion, lower values for $\sigma_{\rm cl}$ and $\zeta_{\rm cl}$
are favored. Giving equal weight to all multiply imaged objects, the
inner multiple image systems are fitted better than the outer ones,
because a high weight is given to the central mass profile. With an
increased weight to the outer images, the outer positions can be 
fitted better, at the
cost of the central images. This is expected since we have more
observables (12 image positions = 24 observables) and up to 11 to 13
free parameters (4 for the cluster, 3 for the cD, one for DB, three
for the redshifts of the W, C and B system, possibly also one for V
and S) and one can not expect the true mass distribution to follow the
model in every detail.  Most likely, there is a change of the
ellipticity and orientation of the cluster potential with radius as it
is visible in the cD light and in the X-ray contours.

The best fit cluster velocity dispersions are of the order of $600
{\rm kms/s}$ and thus are at the low end of the dynamical estimates of
Carlberg et al. (1996a, b). However, the amplitude $\psi_0 \propto
\sigma_{\rm cl}^2$ determined by the lens modelling can be easily
related to the `real' dynamical velocity dispersion only in the
spherical symmetric case where $\sigma_{\rm cl} \equiv \sigma_{\rm
dyn}$ should hold.  Under the two assumptions, the weak lensing
analysis confirms a low velocity dispersion $\sigma_{\rm cl}$ of the
cluster: The first is that the core radius of the cluster is
approximately $7\farcs 5$; this assumption is assisted by the X-ray
results of Hamana et al. (1997) [$(6.9\pm 1.4)\arcsec$ or $(7.5 \pm
1.5) \arcsec$, according to the two cases analyzed there] which
normally do not overestimate the core radius of the dark mass
profile. And secondly we have assumed -- after comparing the number
counts to that of the I-band and using the CFRS-results-- that the
mean redshift of the galaxies used for the weak lensing analysis is
one. Since the cluster velocity dispersion is in agreement with
the value of Carlberg et al. (1996 a\&b), this is also the case for
the total mass estimate and the mass to light ratio of the cluster: i)
if we consider only the velocity dispersions $\sigma_{\rm cl}\approx
600 {\rm km/s} $ and $\sigma_{\rm dyn}=(690\pm 100){\rm km/s}$ derived
from lensing and peculiar velocities and ignore the asymmetry for the
mass estimate within the virial radius, the lensing mass is $75\%$ of
the dynamical mass, but of course compatible to it within the error
bars.  ii) Including all galaxies considered for the light deflection,
we obtain (for a Einstein-de Sitter cosmology) a mass within chip-3 of
$\approx 8*10^{13} h_{\rm 50}^{-1} M_\o$, which transfers to a mass of
$1*10^{15}h_{\rm 50}^{-1} M_\o $ within the virial radius of $1.803
h_{\rm 100 }^{-1}{\rm Mpc}$ (Carlberg et al. 1996b), if the combined
mass profile of the cluster and its galaxies is isothermal out to the
virial radius.  The Carlberg et al. (1996a) value is $5.5*10^{14}
h_{\rm 100}^{-1} M_\o $ (for $\Omega=0.2$) and this estimate decreases
by about 10 percent if one uses $\Omega=1$.

Hamana et al. (1997) suggested to use the measured quantities (in
their case the X-ray temperature and the core radius of MS1512) to
model the lensing of cB58 and to derive limits for the cosmological
constant (for flat universes). Although velocity dispersion estimates
would drop by $\approx 15\%$ for a flat $\Omega=0.3$ universe relative
to an Einstein-de Sitter universe, we do not consider this as a
promising method, because the true mass distribution is only described
approximately by two elliptical potentials, and the best fit value
$\sigma_{\rm cl}$ can not straightforwardly be related to the measured
$\sigma_{\rm dyn}$. A comparison of the relative lensing strengths for
the multiple images at different source redshifts will not improve
this situation, although these estimates are less model dependent. The
reason is that the lensing strength of a $z=3.17$ source relative to
that of cB58 equals $1.0250$, $1.0217$ and $1.0195$ for an Einstein-de
Sitter, a flat $\Omega=0.3$ and a $\Omega=0.3$ universe with vanishing
cosmological constant, and thus the fractional differences are in the
permille range.

Using imaging and spectroscopy under excellent seeing conditions could
further improve the lens models: Deeper photometry in V and R or
additional filters can show that B3 is the third image corresponding
to B1 and B2.  If the flux ratios of all the multiply imaged galaxies
can be obtained more accurately, they can quantitatively be included
into the lens modelling. We consider this as difficult since then a
filter needs to be chosen where the absorption and emission of the cD
are small, and where the blue galaxies C and W are bright enough.  If
the redshift of V (and not so important but much easier: the redshift
of S) is determined, a two-screen model can be used. The
uncertainty in the lensing strength of B and C (see Table 2) is much
larger then the minimum uncertainty given by the unknown cosmological
parameters. Therefore redshift measurements of the C and B-system can
improve the constraint on the slope and thus the core radius of the
potential. 

One of the goals of our investigations was to show that cB58 is a fairly
`normal' high-redshift galaxy. This becomes most obvious when the
counterimage A2 is considered and its magnification of $\mu_{A2}=2-3.5$
is taken into account. Therefore, cB58 is $3.35$-$4$ magnitudes
brighter than its unlensed (total) source. If one shifts the data
point for cB58 in the magnitude-redshift plane in Fig. 5 of Lowenthal
et al. (1997) by that amount, one sees that the source of cB58 is a
normal `$z=3$-Steidel galaxy' in the I-magnitude - redshift
plane. This is also valid for the source size: the half-light radius
of A2 is $0\farcs 25 \pm 0\farcs 05$ in the V and R band, which equals
$3 h_{50}^{-1}{\rm kpc}$, $3.75 h_{50}^{-1}{\rm kpc}$ and $1.9
h_{50}^{-1}{\rm kpc}$ in a $q_0=0.05$, a Einstein-de Sitter, and a
flat universe with $\Omega=0.1$. The half-light angle agrees with that
found by Steidel et al. (1996b) for galaxies of the same redshift.
Including the minimum magnification $\mu_{A2}\ga 2$ of A2 decreases
the half-light radius by 30 percent, i.e. to $0\farcs 18$ or $2.1
h_{50}^{-1} {\rm kpc}$ for $q_0=0.05$.  Thus the unlensed source A
agrees in R-magnitude, redshift and half light radius with the galaxy
C2-05 in the HDF (Steidel et al. 1996b).
To compare with absolute B-band luminosity of other high redshift
galaxies (Lowenthal et al. 1997, Fig. 6) we 
avoid the uncertainties related to the k$_V$ correction and 
simply consider  the  H band which is approximately equivalent
 to the rest B band at z=2.72.
We then correct the $m_H = 19.82$  for extinction according to
the $ E(B-V)$ = 0.3  given by Ellingson et al. (1996).
The authors found that a  10 Myr old constant star formation model
with this amount of extinction provides the best
fit to the optical-IR data of the galaxy. 
Therefore, taking into account the uncertainties in the magnification 
factor and the extinction correction, the m$_H$  implies
a rest frame B absolute magnitude in the range
$ -25.43 \leq$  M$_B$ $\leq -24.05$ ({\it H$_0$} = 75 km$^{-1}$
sec$^{-1}$ Mpc$^{-1}$, {\it q$_0$}=0.05) quite
close to the M$_B \simeq -24$ in  Lowenthal et al. (1997).  An even
better agreement is obtained if the lower value $E(B-V)=0.1$ adopted
by Lowenthal et al. (1997) is used. 

The enormous size and flux of cB58 can be attributed to the gravitational
lens effect. This is of course not the case for the surface
brightness, which is conserved by lensing. Since the surface
brightness stays high all over cB58 and it is fairly structureless
(despite of the spots in Fig. 3, which are not visible from ground) 
this was interpreted as an indication for a 
 simultaneous and high star formation spread over the entire galaxy, and
therefore  cB58 was denoted  a primeval galaxy. 
The reconstruction of the source of the arc shows  that only a part of
the source is lensed into cB58, and that those regions with highest
surface brightness experience the largest magnification. The light
distribution of cB58 is a `zoom' into the central part of its source
where the star formation rate is high. The comparison of the spectral
properties (line width and ratios) of cB58 and its counterimage will
therefore give limits on the inhomogeneity of the star formation in
the source of cB58 with unprecedented spatial resolution.

It seems surprising that a `weak' cluster with a velocity dispersion
of $\sigma\approx 600 {\rm km/s}$ can act as a strong lens, and that
not only one but four multiple image systems are found. This can be
attributed to the high surface density of $z>2$ galaxies (Lowenthal et
al. 1997) and the increased lensing strength for these sources,
compared to $z\approx 1$ galaxies. With the sources A, B, C, W and the
arclet lying $12\arcsec$ N-NE of the cD (which is similar in color and
surface brightness to B1/B2) we have at least 5 candidates with a
redshift larger than $2$ (or $2.5$) within a radius of $15$ arcseconds
around the cluster center, corresponding to a large number density of
$\approx 25$ galaxies per square arcminute. This number density
increases when one takes into account that these sources originate
from a much smaller area in the source plane and that the slope of the
logarithmic number counts at those redshifts is probably not steeper
than $0.4$.  Comparison to predicted and observed high redshift number
counts are difficult due to the individual magnification of the
galaxies and since these investigations use galaxies with a flux limit
in the I or `I+V'-band. The large number of high redshift galaxies can
also be caused by statistical fluctuations or by a group of galaxies
at $z=2.7$, since we can not exclude that the sources of the W and
B-system are at the same redshift as cB58. Our investigation show as
well as that of Franx et al. (1997) and Trager et al. (1997), that the
analysis of galaxies lensed by foreground clusters provides a highly
successful method to find high-redshift galaxies.

Finally, we have shown that a cluster with a velocity dispersion as
low as $\sigma\approx 600{\rm km/s}$ is not only detectable by weak
lensing methods, but its velocity dispersion is still measurable
within an accuracy of $150{\rm km/s}$ at a one sigma level. This
confirms the claim (Schneider 1996) that even less massive dark matter
halos can be detected at a statistically significant level under the
same observing conditions as here and that halos of the same depth can
be detected in shallower observations with a larger point spread
function.

%\vfill\eject 
\section*{Acknowledgments}
We thank J.-P. Kneib and Y. Mellier for discussion on the multiple
image systems. We are grateful to G. Luppino for providing us with the
ground-based data on MS1512 in the R- and B-band and to R. Carlberg
for providing us with the positions of redshift-confirmed cluster
members. We also thank P. Schneider for encouraging discussion and
valuable suggestions on the manuscript.  This work was supported by
the ``Sonderforschungsbereich 375-95 f\"ur
Astro-Teilchenphysik'' der Deutschen Forschungsgemeinschaft.


\begin{thebibliography}{99}
%
\bibitem{ref:1} %angul-diam dist and cosm param,dds/ds ratios ...
Asada, H., 1996, astro-ph/9611110
\bibitem{ref:2}
Bender, R., Saglia, R.P., Ziegler, B., Belloni, P., Greggio, L., Hopp,
U., Bruzual, G., 1997, submitted to ApJL
%
\bibitem{ref:3}
Bender, R., M\"ollenhoff, C., 1987, A\&A, 177, 71
%
\bibitem{ref:4}
Bertin, E., Arnouts, S., 1996 , A\&AS, 117, 393
%
\bibitem{ref:5}
Bruzual, G.A., Charlot, S., 1993, ApJ, 405, 538
%
\bibitem{ref:6}
Bruzual, G.A., Charlot, S., 1997, submitted to ApJ
%
\bibitem{ref:7} % veloc disp
Carlberg, R., Yee, H.K.C., Ellingson, E., Abraham, R., Gravel, P.,
Morris, S., Pritchet, C.J., 1996a, ApJ, 462, 32 
%
\bibitem{ref:8} % VLT work shop		
Carlberg, R., Yee, H.K.C., Ellingson, E., Morris, S., Abraham, R.,
 Gravel, P., Hartwick, F.D.A., Hesser, J.E., Hutchings, J.B., Oke,
J.B., Pritchet, C.J., Smecker-Hane, T., 1996b, ApJ, 462, 32
% 
\bibitem{ref:9} %D_n-sigma relation
Dressler, A., Faber, S.M., Burstein, D., Davies, R.L., Lynden-Bell,
D., Terlevich, R.J., Wegner, G., 1987, ApJ, 313, L37
%
\bibitem{ref:10} % redshifts of arclets ?
Ebbels, T.M.D., Le Borgne, J.F., Pello, R., Ellis, R.S., Kneib, J.-P., 
Smail, I., Sanahuja, B., 1996, MNRAS, 281L, 75
%
\bibitem{ref:11}
Ellingson, E., Yee, H.K.C., Bechtold, J., Elston, R., 1997, ApJ, 466, L71
%
%\bibitem{ref:12}
%Fort, B., Mellier, Y., 1994, A\&AR, 5,  239
%
\bibitem{ref:13}
Franx, M., Illingworth, G.D., Kelson, D.D., van Dokkum, P.G., Tran,
K-V, 1997, preprint
%
\bibitem{ref:14}
Frayer, D.T., Papadopoulos, P.P., Bechthold, J., Seaquist, E.R., Yee,
H.K.C., 1997, ApJ, 113, 562
%
\bibitem{ref:15} %angul-diam dist
Fukugita, M., Futamase, T., Kasai, M., Turner, E.L., 1992, ApJ, 393, 3
%
\bibitem{ref:16}
Gioia, I.M., Luppino, G.A., 1994, ApJS, 94, 583.
%
\bibitem{ref:17} %xray-stuff
Hamana, T., Hattori, M., Ebeling, H., Henry, J.P., Futamase, T.,
Shioya, Y., 1997, astro-ph/9703136
%
\bibitem{ref:18}
Holtzman, J.A., Burrows, C.J., Casertano, S., Hester, J.J., Trauger,
J.T., Watson, A.M., Worthey, G., 1995, PASP, 107, 1065
%
\bibitem{ref:19}  %A370
Kneib, J.-P., Mellier, Y., Fort, B.,  Mathez, G., 1993, A\&A, 273, 367
%
\bibitem{ref:20} %A2218, gb
Kneib, J.-P., Mathez, G., Fort, B., Mellier, Y., Soucail, G.,
Longaretti, P.-Y., 1994, A\&A, 286, 701
%
\bibitem{ref:21} %A2218, hst
Kneib, J.-P., Ellis, R.S., Smail, I., Couch, W.J., Sharples, R.M.,
1996, ApJ, 471, 643
%
\bibitem{ref:22} %CFRS-pred for faint I-magnitudes
Lilly, S.J., Tresse, L., Hammer, F., Crampton, D., Le Fevre, O., 1995,
ApJ, 455, 108
%
\bibitem{ref:23} % U and B `dropouts'
Lowenthal, J.D., Koo, D.C.,  Guzman, R., Gallego, J., Phillips, A.C.,
Faber, S.M.,  Vogt, N.P., Illingworth, G.D., Gronwall, C., 1997, ApJ,
481, L673
%
\bibitem{ref:24} % LR-deconvolution
Lucy, L., 1974, AJ, 79, 745
%
\bibitem{ref:25}  %A370
Mellier, Y., Soucail, G., Fort, B., Le-Borgne, J.-F., Pello, R., 1990,
in {\it Gravitational Lensing}, Mellier, Y., Fort, B., Soucail,
G. (eds.), Springer
%
\bibitem{ref:26} %MS 2137
Mellier, Y., Fort, B., Kneib, J.-P., 1993, ApJ, 407, 33 
%
\bibitem{ref:27} % numerical recipes
Press, W.H., Teukolsky, S.A., Vettering, W.T. \& Flannery,
B.P., 1992, {\it Numerical Recipes}, Cambridge University Press
%
\bibitem{ref:28}
Saglia, R.P., Bender, R.,  Ziegler, B., Belloni, P., Greggio, L., Hopp,
U., Bruzual, G., 1997, in preparation
%
\bibitem{ref:29}
Schneider, P., Ehlers, J., Falco, E., 1992, {\it Gravitational
Lenses}, Springer, Heidelberg
%
\bibitem{ref:30}
Schneider, P., Seitz, C.,  1995, A\&A, 294, 411 %  \chi definition
% 
\bibitem{ref:31} % dark matter halos
Schneider, P., 1996, A\&A, 283, 837
\bibitem{ref:32}
% Cl0939
Seitz, C., Kneib, J.P., Schneider, P.,  Seitz, S., 1996 A\&A,
314, 707
%
\bibitem{ref:33}  %Number counts
Smail, I., Hogg, D.W., Yan, L. \& Cohen, J.G., 1995, ApJ, 449,
L105
%
\bibitem{ref:34}
Steidel, C.C., Giavalisco, M., Pettini, M., Dickinson, M., Adelberger,
K.L., 1996a, ApJ, 462, 17
%
\bibitem{ref:35}
Steidel, C.C., Giavalisco, M.,  M., Dickinson, M., Adelberger,
K.L., 1996b, AJ, 112, 352
%
\bibitem{ref:36}
Trager, S.C., Faber, S.M., Dressler, A., Oemler, A.Jr., 1997, astro-ph/9703062
%
\bibitem{ref:37}
Williams, L. L. R.,  Lewis, G.F., 1997, MNRAS, 281, L35 (WL)
%
\bibitem{ref:38}
 Yee, H. K. C., Ellingson, E., Bechtold, J., Carlberg, R. G.,
Cuillandre, J.-C., 1996, AJ, 111, 1883 (Y96)
%
\bibitem{ref:39}
Ziegler, B., Bender, R., 1997, MNRAS, in press (astro-ph/9704280)
%
\bibitem{ref:40}
Ziegler, B., 1996, PhD Thesis, University Heidelberg
%
\bibitem{ref:41}
Ziegler, B., 1997, in preparation
%
\end{thebibliography}
\end{document}